\documentclass[11pt, letterpaper]{article}

\RequirePackage{doi}

\usepackage{amsmath}
\usepackage{amssymb}
\usepackage{amsthm}
\usepackage{graphicx}
\usepackage{hyperref}
\usepackage{url}
\usepackage{algorithm}
\usepackage{algorithmic}
\usepackage{caption}
\usepackage{subcaption}
\usepackage{booktabs}
\usepackage{authblk}
\usepackage{appendix}
\usepackage[utf8]{inputenc}
\usepackage[T1]{fontenc}
\usepackage[english]{babel}
\usepackage{geometry}
\usepackage{xcolor}
\usepackage{fancyhdr}
\usepackage{microtype}
\usepackage{parskip}
\usepackage{tabularx}
\usepackage{enumitem}

\hypersetup{
    colorlinks=true,
    linkcolor=blue,
    filecolor=magenta,      
    urlcolor=cyan,
    citecolor=red,
}

\definecolor{headercolor}{RGB}{0, 50, 100}
\title{Large Language Models for Bioinformatics}
\date{January 2025}

\newcommand*\samethanks[1][\value{footnote}]{\footnotemark[#1]}

\author[1]{Wei Ruan\thanks{Co-first authors.}}
\author[2]{Yanjun Lyu\samethanks}
\author[2]{Jing Zhang\samethanks}
\author[3]{Jiazhang Cai}
\author[1]{Peng Shu}
\author[4]{Yang Ge}
\author[4]{Yao Lu}
\author[5]{Shang Gao}
\author[1]{Yue Wang}
\author[6]{Peilong Wang}
\author[1]{Lin Zhao}
\author[3]{Tao Wang}
\author[3]{Yufang Liu}
\author[3]{Luyang Fang}
\author[3]{Ziyu Liu}
\author[1]{Zhengliang Liu}
\author[1]{Yiwei Li}
\author[1]{Zihao Wu}
\author[1]{Junhao Chen}
\author[1]{Hanqi Jiang}
\author[1]{Yi Pan}
\author[1]{Zhenyuan Yang}
\author[6]{Jingyuan Chen}
\author[7]{Shizhe Liang}
\author[8]{Wei Zhang}
\author[9]{Terry Ma}
\author[10]{Yuan Dou}
\author[10]{Jianli Zhang}
\author[10]{Xinyu Gong}
\author[10]{Qi Gan}
\author[10]{Yusong Zou}
\author[10]{Zebang Chen}
\author[10]{Yuanxin Qian}
\author[10]{Shuo Yu}
\author[1]{Jin Lu}
\author[10]{Kenan Song}
\author[10]{Xianqiao Wang}
\author[11]{Andrea Sikora}
\author[12]{Gang Li}
\author[13]{Xiang Li}
\author[13]{Quanzheng Li}
\author[14]{Yingfeng Wang}
\author[15]{Lu Zhang}
\author[16]{Yohannes Abate}
\author[17]{Lifang He}
\author[3]{Wenxuan Zhong}
\author[3]{Rongjie Liu}
\author[4]{Chao Huang}
\author[6]{Wei Liu}
\author[4]{Ye Shen}
\author[3]{Ping Ma}
\author[5]{Hongtu Zhu}
\author[10]{Yajun Yan}
\author[2]{Dajiang Zhu\thanks{Corresponding authors.}}
\author[1]{Tianming Liu\samethanks}

\affil[1]{School of Computing, University of Georgia, GA, USA}
\affil[2]{Department of Computer Science and Engineering, University of Texas at Arlington, TX,
USA}
\affil[3]{
Department of Statistics, University of Georgia, GA, USA}
\affil[4]{Department of Epidemiology and Biostatistics, University of Georgia, Athens, GA, USA}
\affil[5]{Department of Biostatistics, UNC Chapel Hill, NC, USA}
\affil[6]{Department of Radiation Oncology, Mayo Clinic, Phoenix, AZ, USA}
\affil[7]{Institute of Plant Breeding, Genetics \& Genomics, University of Georgia, Athens, GA, USA}
\affil[8]{School of Computer and Cyber Sciences, Augusta University, GA, USA}
\affil[9]{School of Computer Science, Carnegie Mellon University, Pittsburgh, PA, USA}
\affil[10]{College of Engineering, University of Georgia, Athens, GA, USA}
\affil[11]{Department of Biomedical Informatics, University of Colorado, CO, USA}
\affil[12]{Department of Radiology, University of North Carolina at Chapel Hill, NC, USA}
\affil[13]{Department of Radiology, Massachusetts General Hospital and Harvard Medical School, MA, USA}
\affil[14]{Department of Computer Science and Engineering, University of Tennessee at Chattanooga, TN, USA}
\affil[15]{Department of Computer Science, Indiana University Indianapolis, IN, USA}
\affil[16]{Department of Physics and Astronomy, University of Georgia, Athens, GA, USA}

\affil[17]{Department of Computer Science and Engineering, Lehigh University, PA, USA}

\begin{document}

\maketitle

\begin{abstract}

With the rapid advancements in large language model (LLM) technology and the emergence of bioinformatics-specific language models (BioLMs), there is a growing need for a comprehensive analysis of the current landscape, computational characteristics, and diverse applications. This survey aims to address this need by providing a thorough review of BioLMs, focusing on their evolution, classification, and distinguishing features, alongside a detailed examination of training methodologies, datasets, and evaluation frameworks. We explore the wide-ranging applications of BioLMs in critical areas such as disease diagnosis, drug discovery, and vaccine development, highlighting their impact and transformative potential in bioinformatics. We identify key challenges and limitations inherent in BioLMs, including data privacy and security concerns, interpretability issues, biases in training data and model outputs, and domain adaptation complexities. Finally, we highlight emerging trends and future directions, offering valuable insights to guide researchers and clinicians toward advancing BioLMs for increasingly sophisticated biological and clinical applications.

\end{abstract}

\section{Introduction}

The rapid development of large language models (LLMs)  such as BERT~\cite{kenton2019bert}, GPT~\cite{brown2020language}, and their specialized counterparts has revolutionized the field of natural language processing (NLP). Their ability to model context, interpret complex data patterns, and generate human-like responses has naturally extended their applicability to bioinformatics, where biological sequences often mirror the structure and complexity of human languages~\cite{liu2024large}. LLMs have been successfully applied across various bioinformatics domains, including genomics, proteomics, and drug discovery, offering insights that were previously unattainable through traditional computational methods~\cite{sarumi2024large}.

Despite significant advancements, challenges remain in the systematic categorization and comprehensive evaluation of applications of these models on bioinformatic problems. Considering the variety of bioinformatics data and the complexity of life activities, navigating the field can often be challenging, as existing studies tend to focus on a limited scope of applications. This leaves gaps in understanding the broader utility of LLMs in various bioinformatics subfields~\cite{tripathi2024large}.

This survey aims to address these challenges by providing a comprehensive overview of LLM applications in bioinformatics. 
By focusing on different levels of life activities, this article collected and exhibited related works from two major views: life science and biomedical applications.

We have collaborated with domain experts to compile a thorough analysis spanning key areas in these views, such as nucleoid analysis, protein structure and function prediction, genomics, drug discovery, and disease modeling, including applications in brain diseases and cancers, as well as vaccine development. 

In addition, we propose the new term `Life Active Factors' (LAFs) to describe the molecular and cellular components that serve as candidates for life science research targets, which widely includes not only concrete entities (DNA, RNA, protein, genes, drugs) but also abstract components (bio-pathways, regulators, gene-networks, protein interactions) and biological measurements (phenotypes, disease biomarkers). LAFs is a comprehensive term that is capable of reconciling the conceptual divergence arising from research across various bioinformatics subfields, benefiting the understanding of multi-modality data for LAFs and their interplays in complex bio-systems. The introduction of LAFs aligns well with the spirit of foundational models and emphasizes the unification across sequence, structure, and function of the LAFs while respecting the interrelationships of each LAF as a node within the biological network.

\begin{figure}[!h]
    \centering
    \includegraphics[width=0.99\textwidth]{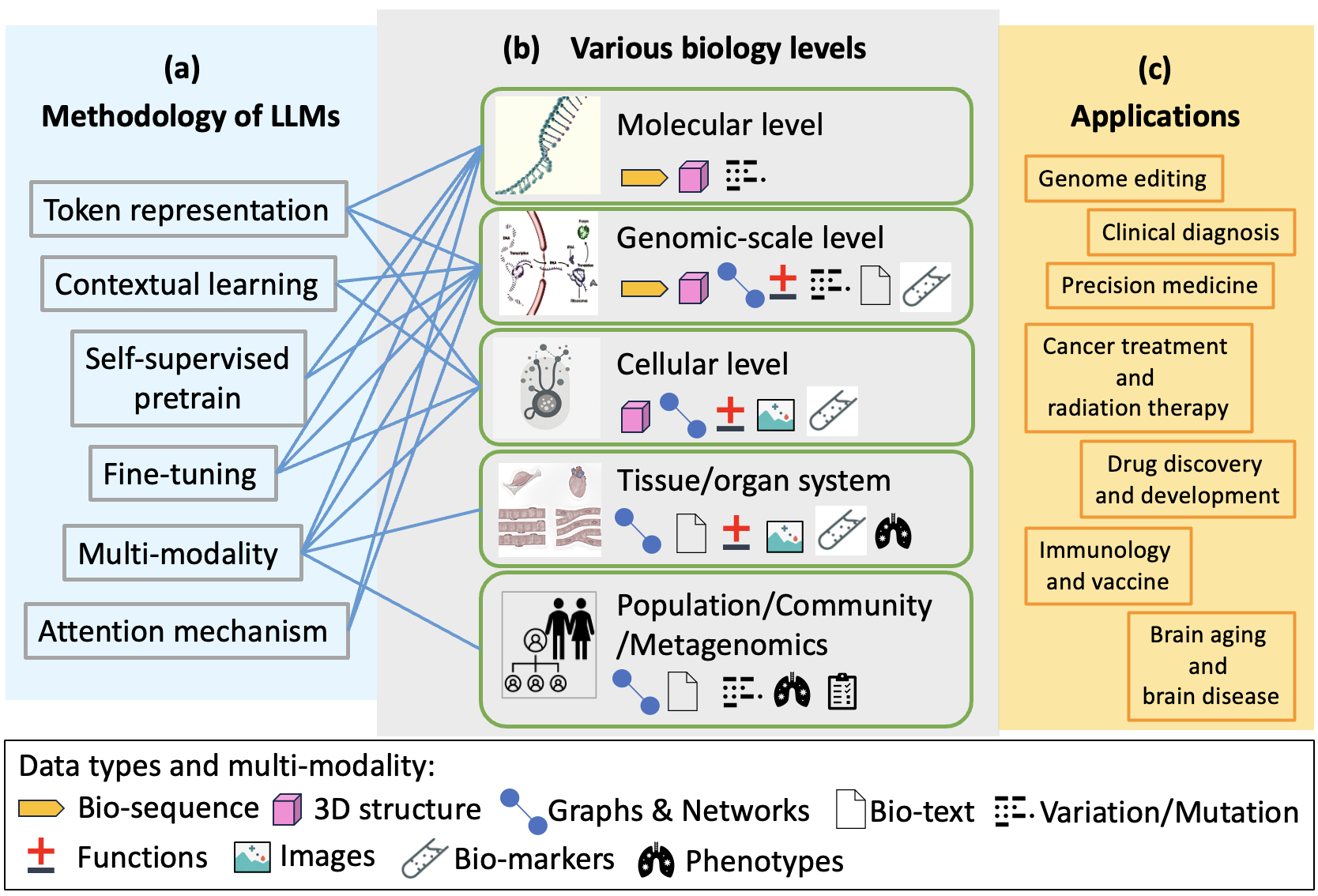}
    \caption{The applications of the methodology of LLMs in bioinformatics tasks.}
    \label{fig:main}
\end{figure}

By bridging existing knowledge gaps, this work seeks to equip bioinformaticians, biologists, clinicians, and computational researchers with an understanding of how LLMs can be effectively leveraged to tackle pressing problems in bioinformatics. Our survey not only highlights recent advances but also identifies open challenges and opportunities, laying the foundation for future interdisciplinary collaboration and innovation (Figure ~\ref{fig:main}).

\section{Background of Language Models and Foundation Models in Bioinformatics}

Bioinformatics has become a fundamental and transformative field in life sciences, bridging computational techniques and biological research. It emphasizes the development and application of computational tools and methodologies to manage and interpret vast amounts of biomedical data, transforming them into actionable insights and driving advancements across diverse downstream applications. Modern computational tools, particularly those rooted in deep learning technology, have significantly accelerated the evolution of biological research.

The rapid advancements in LLMs technologies have inspired new approaches to bioinformatics computing. Considering the complexity of biological systems and highly structured nature of bioinformatics data, LLM-based computing methods have proven effective in addressing challenges across fields such as genomics, proteomics, and molecular biology. Inspired by LLM architectures like transformers, foundation models in bioinformatics excel at capturing complex patterns and relationships in biological data. They have evolved from single-modality tools to sophisticated multimodal systems, integrating diverse datasets such as genomic sequences and protein structures.

Central to their success is the availability of large-scale, high-quality training data and the adoption of self-supervised pretraining and fine-tuning techniques. These methods allow models to extract meaningful features from unlabeled data and adapt to specific bioinformatics tasks. Together with advances in architecture design, these innovations have broadened the capabilities and impact of foundation models, unlocking new insights into biological systems and accelerating progress in the life sciences. The following sections discuss these advanced computing methods along with the intrinsic properties of biological systems and structured bioinformatics data.

\subsection{Foundations of Language Models and Bioinformatics Overview}
\subsubsection{Basics of Large Language Models and Foundations Models}
Traditional language models are engineered to process and generate text in a human-like manner, leveraging the extensive datasets used during their training. These models excel at interpreting context, producing coherent and contextually appropriate responses, performing translations, summarizing text, and answering questions. LLMs are a type of foundation model trained on vast datasets to provide flexible and powerful capabilities~\cite{liu2023summary,zhao2023brain,zhong2024evaluation} that address a broad spectrum of use cases and applications~\cite{ma2024iterative,dai2023chataug,liu2023radiology,liao2023differentiating,liu2023context,rezayi2022clinicalradiobert,dai2023ad,zhao2023ophtha,zhang2024generalist,liu2023radonc,liu2024fine,lyu2024gp,wang2024comprehensive,huang2024position,liu2024surviving,huang2024trustllm,zhenyuan2024analyzing,huang2024position,wang2023prompt,liu2023tailoring,tian2024assessing,lee2023multimodality,shu2024llms,latif2023artificial,wang2024large,liu2024understanding,latif2024systematic,li2024artificial,zhong2024evaluation,wang2024recent,ding2024foundation,chen2024queen,zhang2024potential,zhong2024opportunities,jiang2024oraclesage,liao2024zero,zhang2024generalizable,tan2023promises,shi2024mgh,shu2024transcending,wang2024legal,holmes2024benchmarking,zhao2022embedding,zhao2023generic,li2024artificial,lyu2024gp,liu2024llm,yang2024examining,wang2024comprehensive,liu2024fine,gong2024advancing,mukherjee2024polaris,xu2024reasoning,latif2024knowledge,liu2024radiation,huang2024trustllm,wei2023chat2brain,liu2023holistic,liu2022survey,wu2023exploring,xiao2023instruction,cai2023exploring,holmes2023evaluating,liu2023pharmacygpt,guan2023cohortgpt,liu2023evaluating,cai2023multimodal,shi2023mededit,tang2023policygpt,liu2023transformation,zhong2023chatradio,gong2023evaluating,liao2023coarse,holmes2023evaluatingbio,rezayi2024exploring,dou2023towards,holmes2023benchmarking}. By efficiently handling diverse tasks, LLMs eliminate the need for building and training separate domain-specific models for each use case—a process that is often limited by cost and resource constraints. This unified approach not only fosters synergies across tasks but also frequently results in superior performance, making LLMs a more scalable and efficient solution. There are several key elements that make the language model successful in adaptation to bioinformatics tasks (Figure ~\ref{fig:main}(a)).

\textbf{Representation learning and tokenization}
Tokenization in LLMs is influenced by the design of their tokenization algorithms, which primarily use subword-level vocabularies to represent text sequence data effectively. Popular tokenization algorithms, such as Byte-Pair Encoding (BPE)~\cite{sennrich2015neural}, WordPiece~\cite{schuster2012japanese}, and Unigram~\cite{kudo2018subword}, are widely used. Although their vocabularies cannot perfectly capture every possible variation of input expressions, these tokenization methods effectively encode the features of words and their contextual relationships.

In the view of representation learning, the tokenization and token embedding algorithms of the language model generally succeeded in representing the hidden factors of variation behind the data. This representation is based on the unsupervised learning scheme of the language models. The sub-word context features learned in the encoder modules or embedding layers follow the probabilistic modeling and continuously update the representations on large corpus datasets~\cite{bengio2013representation}.
 
\textbf{Attention mechanism}
LLMs widely use the transformer model~\cite{vaswani2017attention, lin2017structured} as their foundational architecture. A core innovation of the transformer model is the multi-head self-attention mechanism, which establishes relationships among all relevant tokens, enabling more effective encoding of each word in the input sequence. The self-attention layer processes a sequence of tokens (analogous to words in a language) and learns context information across the entire sequence. The "multi-head" aspect refers to multiple attention heads operating simultaneously to capture diverse contextual features. Inside a single attention head, a token output embedding in a sequence is computed and fused with other tokens in the context with a proper causal mask. Such global level attention mechanic enables efficient information fusion along available context windows.

\textbf{Self-supervised training methods}
Language models are trained using self-supervised learning methods~\cite{liu2023self}. Unlike supervised learning, which typically requires human annotations, language models can leverage vast amounts of unannotated text data~\cite{hastie2009overview}. The objective of unsupervised learning is to analyze unlabeled data by identifying and capturing its meaningful properties. Neural networks can extend some of these approaches. For example, autoencoders compress data into a low-dimensional representation through a hidden layer known as the bottleneck layer and then reconstruct the original input data from this representation~\cite{hinton2006reducing, kramer1991nonlinear, vincent2008extracting, vincent2010stacked}.
Language models leverage either the next word in a sentence as a natural label for the context or artificially mask a known word and predict it. This method, where unstructured data generates its own labels (e.g., predicting the next word or a masked word) and language models are trained to predict them, is known as self-supervised learning. Transformer-based models, with their parallel processing capabilities and ability to capture correlations across entire sequences, have achieved state-of-the-art (SOTA) performance ~\cite{petroni2019language, howard2018universal}. A more advanced training diagram is the text-to-text framework. This kind of training diagram unified multiple kinds of tasks, including translation, question answering, classification, formulated and feeding to model as input and training it as a generative model to predict target text. This framework, which is named ‘T5’ benefits using the same model, loss function, hyperparameters, etc. across a diverse set of tasks~\cite{raffel2020exploring}.

\textbf{Pre-training}
In many supervised learning problems, input data is represented by multiple features, comprising numerical or categorical information that can aid in making predictions. Scratch-trained models, which initialize and train all parameters from the ground up using task-specific datasets, typically require numerous iterations to converge fully on a single task. In general, transformer-based language models fall into two categories: scratch-trained models and pre-trained models. LLMs apply transformer-based pre-trained models that are trained from large amounts of unlabeled data and then fine-tuned for specific tasks. Pre-training learns general information from unlabeled data which can improve the convergence rate of the target tasks and often  has better generalization than training parameters from scratch~\cite{han2021pre}. The use of context information in a large corpus to pre-train the whole model (or encoder modules) has achieved SOTA results in various downstream tasks.

\subsubsection{Bioinformatics Applications and Challenges}
Using deep learning methods like language models to tackle bioinformatics problems is challenging. While deep learning models have shown superior accuracy in specific bioinformatics applications (e.g., genomics applications) compared to SOTA approaches and are adept at handling multimodal and highly heterogeneous data, significant challenges remain. Further work is required to integrate and analyze diverse datasets required for deep learning for genomic prediction and prognostic tasks. This is especially important for the development of explainable language models that can identify novel biomarkers and elucidate regulatory interactions across various biology levels: pathological conditions, including different tissues and disease states. These advancements require a deep understanding of complex bioinformatics data, the related tasks, and their mutual relationships~\cite{koumakis2020deep}.
In this review, we discuss such issues through two lens: the various biology levels and the inherent regulations of life activities.

\textbf{Various biology levels}
Although no gold standard division was available, the levels of life-science factors in bioinformatics can be divided into five levels, from micro to macro. Here, we take the mammal model organisms as a template, the levels can be divided into: the molecular level, the genome-scale level, the cellular level, the tissue/organ system level, and the population/community/metagenomics level (Figure ~\ref{fig:main}(b)). Bioinformatics often focuses on the first three levels (i.e., the molecular level, the genomic-scale level, and the cellular level). The molecular level analysis targets involved biologically active molecules, which include nucleic acids, amino acids, and other small bioactive molecules, and the relative experiments aimed at interpreting the life activities at this scale. The genomic-scale level models the life activities from DNA, RNA, and proteins to metabolomics. The most famous regulation at the genomic scale level is The Central Dogma, which reveals the intrinsic relations of main life-activity factors on a sub-cellular scale. The whole sub-cellular system is modeled hierarchically, beginning with DNA, mRNA, and proteins, extending to metabolomics, and ultimately inferring the phenotype~\cite{watson1958protein}.
At the cellular level, understanding cellular mechanisms is a fundamental challenge in biology and holds significant importance in biomedical fields, particularly concerning disease phenotypes and precision medicine. Using Genes (The specific sequences of nucleotides within DNA that control downstream life activities) as a unit, the functions of genes and the gene products are essential research targets at this level. A comprehensive, structured, computation-accessible representation of gene function and variations is crucial for bioinformatic understanding of the cellular organism or virus. At the same time, the gene networks and mutual influences of gene products pose a challenge for such areas. 
Single-cell sequencing technologies allow us to obtain gene expression data at the mRNA level, providing a foundation for analyzing entire cellular systems. This data is now extensively used to identify cell states during development, characterize specific tissues or organs, and evaluate patient-specific drug responses.
In this review, the molecular components at the genomics level and cellular levels and their respective sets are collectively referred to as Life Active Factors (LAFs). It is important to note that the sequence representation format is the most commonly observed for each LAF. However, multi-modality data for LAF is also significant for representing the property of LAF, i.e., the highly structured data format to record the function descriptions, abundancy, variations, and expressions~\cite{schlitt2003gene, kim2022humannet}.

\textbf{Inherent regulations of life activities} Since most LAFs at each biological level are represented in a sequence format, transformer-based pre-trained language models are particularly well-suited for analyzing these sequences. An emerging consensus suggests that these sequences embody an underlying language that can be deciphered using language models. However, in order to play the roles in life activities, an essential logic of a single LAF is ‘sequences-structures-functions’. Take proteomics analysis as an example, protein sequences can be viewed as a concatenation of letters from the amino acids, analogously to human languages. The latest protein language models utilize these formatted letter representations of secondary structural elements, which combine to form domains responsible for specific functions. The protein language models also direct inference of full atomic-level protein structure from primary sequence and produce functional proteins that evolution would require hundreds of millions of years to uncover~\cite{ofer2021language, lin2023evolutionary, hayes2024simulating}.

In life activities, there are important regulation relationships among the LAFs across different levels as well as intra-level relationships. Considering the genomics level, genes control hereditary traits primarily by regulating the production of RNA and protein products. According to the central dogma of molecular biology, genes within DNA are transcribed into messenger RNA (mRNA), which is then translated into gene products, such as proteins. For any given gene product, whether RNA or protein, its origin can be traced back to the gene that directed its synthesis. This traceability highlights that fully understanding a gene's functionality requires considering not only the gene itself but also the roles and functions of all its associated products.
Genes regulate each other and create feedback loops to form cyclic chains of dependencies in gene regulatory networks, graph neural network-styled operations are suitable to model the “steady state” of genes. It is the same for proteins in protein-protein
interactions (PPI). In the layer of pathways, it is a hypergraph where each hyperedge is a pathway including multiple proteins.

Within the cellular level, pathways integrate individual genes or protein products to perform certain cell functions under mutual intra-level regulations. Proteins interact with one another in various ways, such as inhibiting, activating, or combining with others, thereby influencing expression levels or protein abundances within cells. These interactions are collectively referred to as PPI. Some databases systematically organize results by annotating functionalities using Gene Ontology (GO), utilizing the unique gene identifiers assigned to each gene within the genomic system~\cite{lynch2011transposon, gene2004gene}.


\subsection{Training Methods and Models}


\textbf{Pre-training} is a critical phase in the development of LLMs, where a model learns foundational linguistic representations by training on extensive and diverse datasets. This process typically employs self-supervised learning techniques such as masked language modeling (e.g., BERT~\cite{kenton2019bert}) or causal language modeling (e.g., GPT~\cite{brown2020language}), enabling the model to predict masked tokens or the next word in a sequence. Unlike traditional deep neural networks (DNNs)~\cite{samek2021explaining}, which are often pre-trained on domain-specific datasets such as ImageNet~\cite{deng2009imagenet}, pre-training of LLMs is conducted on significantly larger datasets comprising diverse domains, including books, encyclopedias, and web content. Moreover, pre-training LLMs involves models with billions or even trillions of parameters, making it computationally and resource-intensive compared to conventional DNNs.

The primary advantage of pre-training lies in the model's ability to generalize across diverse language tasks, often achieving zero-shot~\cite{pourpanah2022review} or few-shot~\cite{wang2020generalizing} performance without additional task-specific training. This broad generalization enables LLMs to excel in tasks spanning natural language understanding, generation, and reasoning. However, the disadvantages of pre-training include high computational and energy costs, often requiring distributed systems with high-performance hardware. Additionally, pre-trained models can inherit biases and errors present in the training corpus~\cite{navigli2023biases}, potentially leading to biased or undesirable outputs.

\textbf{Fine-tuning} is the subsequent stage that builds upon the pre-trained model by adapting it to specific tasks or domains through additional supervised or semi-supervised training. This process utilizes smaller, targeted datasets and optimizes the model for a specific use case. Fine-tuning can be categorized into task-specific fine-tuning~\cite{li2023task,zheng2024fine}, where models are specialized for particular tasks such as sentiment analysis or machine translation; domain-specific fine-tuning~\cite{liu2023radiology, chen2024queen}, which refines the model for specialized fields such as medicine or law; and instruction fine-tuning~\cite{xiao2023instruction,liu2024visual}, where the model is trained to respond to natural language prompts in an aligned manner. Recent advancements in parameter-efficient fine-tuning methods~\cite{ding2023parameter}, such as LoRA (Low-Rank Adaptation)~\cite{hu2021lora} and adapters~\cite{he2021effectiveness}, have further improved the efficiency of this process by updating only a subset of the model's parameters while maintaining the computational benefits of the pre-trained foundation.

Fine-tuning enhances the model’s performance on specific tasks by leveraging domain- or task-specific data, achieving state-of-the-art results in various applications. However, it introduces challenges such as the risk of overfitting to the fine-tuning dataset, potentially diminishing the model’s generalization capabilities. Furthermore, fine-tuning requires high-quality labeled data to ensure reliability and accuracy in specialized applications.

\textbf{Reinforcement Learning with Human Feedback (RLHF)}~\cite{christiano2017deep} represents a crucial additional stage in the training pipeline of large language models, designed to align model outputs with human preferences and expectations. While pre-training and fine-tuning equip the model with general linguistic understanding and task-specific expertise, RLHF optimizes the model’s behavior to produce responses that are more aligned with human values, instructions, or conversational styles, which is particularly critical for applications such as conversational agents, where user interaction quality is paramount.

RLHF involves three primary components: a reward model trained on human-labeled preferences, a reinforcement learning (RL) algorithm to optimize the model's behavior based on the reward model, and iterative human feedback to refine the reward system. The reward model is typically developed by collecting a dataset of model outputs ranked by human evaluators. This ranking serves as the ground truth to train the reward model, which predicts the desirability of a given output. Subsequently, reinforcement learning algorithms, such as Proximal Policy Optimization (PPO)~\cite{schulman2017proximal}, adjust the model parameters to maximize the reward score predicted by the reward model. Besides, Direct Preference Optimization (DPO)~\cite{rafailov2024direct} algorithms operates on a dataset of ranked preferences, directly optimizing the model to prefer the more highly ranked output in each pair.

The primary advantage of RLHF is its ability to align the outputs of a pre-trained and fine-tuned model with human expectations, improving qualities such as coherence, relevance, and ethical compliance. This approach is particularly effective in mitigating undesirable behaviors, such as generating toxic, biased, or irrelevant content. Furthermore, RLHF enables the incorporation of domain-specific human expertise, allowing models to better serve niche applications.
However, RLHF introduces several challenges. First, the quality of human feedback is critical, poorly designed feedback mechanisms or misaligned human preferences can lead to suboptimal or even harmful model behavior. Second, RLHF requires significant resources for human annotation and computationally expensive RL training. Furthermore, over-optimization for the reward model can lead to undesirable artifacts, such as the model exploiting weaknesses in the reward system rather than genuinely improving its outputs—a phenomenon known as "reward hacking"~\cite{amodei2016concrete,pan2024feedback}.

\textbf{Knowledge Distillation.} Knowledge Distillation (KD) has emerged as a key approach for efficient training and deploying LLMs by transferring the knowledge embedded in high-capacity teacher models to smaller, more efficient student models \cite{hinton2015distilling}. In essence, the student model learns to mimic both the predictive outcomes and the internal representation patterns of the teacher, thereby significantly reducing computational costs and memory demands during the pre-training phase \cite{xu2024survey,zhang2023huatuogpt,taori2023stanford}. This methodology promotes the development of leaner LLMs without sacrificing their ability to perform complex language tasks.

Recent advancements in KD extend beyond final-output matching. Modern methods utilize established LLMs to generate not only predictions but also detailed reasoning steps, which are often referred to as chain-of-thought sequences or intermediate logic traces \cite{hsieh2023distilling,liu2024deepseek}. These rich annotations can then be incorporated into the fine-tuning process, enabling the target LLM to acquire deeper problem-solving skills and enhance interpretability without extensive manual labeling. By integrating these reasoning pathways, KD no longer serves solely as a compression mechanism but also imparts advanced critical thinking and inference capabilities to newly trained models.
Moreover, recent work explores expanding KD to support specialized or domain-specific tasks where the established teacher models can guide the target LLM toward focusing on task-relevant knowledge, filtering out less pertinent information \cite{firoozi2023foundation,liu2024ddk}. This approach helps produce models that are better aligned with their intended applications.
Additionally, a Bayesian perspective on KD has been introduced, offering a transparent interpretation of its statistical foundations and equipping the target model with robust uncertainty quantification capabilities \cite{fangbayesian,korattikara2015bayesian}.

The integration of pre-training, fine-tuning, KD, and RLHF represents a comprehensive training paradigm for LLMs. Pre-training serves as the foundation, equipping the model with general knowledge and linguistic capabilities through large-scale unsupervised learning. Fine-tuning adapts the model to specific tasks or domains, enhancing its performance in targeted applications. KD supports efficiency by enabling the transfer of knowledge from established teacher models to target models, while RLHF refines the model's behavior to align with human preferences, ensuring outputs are both functionally accurate and socially acceptable. These stages are complementary and iterative. Insights gained during RLHF can inform improvements in fine-tuning datasets or methodologies, while advancements in fine-tuning and KD can enhance the quality of RLHF outcomes. Together, this pipeline not only ensures that LLMs are powerful and versatile but also makes them more usable and aligned with human-centered goals. This multi-stage training paradigm has been instrumental in the development of state-of-the-art models like OpenAI's ChatGPT and Anthropic's Claude, setting a benchmark for future advancements in the field.
These advancements include the release of both full-scale and lightweight versions, with KD often playing a role in optimizing the latter \cite{touvron2024llama3}.

\subsection{Bioinformatics-Specific Datasets}
The rapid advancements in large language models  have significantly propelled the development of bioinformatics by enabling more efficient data interpretation and knowledge extraction. LLMs excel in understanding, processing, and generating complex textual and numerical data, making them powerful tools for tasks such as sequence analysis, annotation, and predictive modeling~\cite{xu2024towards,tian2024opportunities}. Leveraging bioinformatics-specific datasets, LLMs can further refine their understanding to address domain-specific challenges, transforming raw data into tangible, interpretable forms that accelerate research and innovation.

\textbf{Question Answering (QA)} systems play a vital role in biomedicine, assisting with clinical decision support and powering medical chatbots. The development of robust QA systems relies heavily on diverse and well-curated datasets. Over the past decade, several biomedical QA datasets have been introduced, each targeting specific challenges and domains. For instance,, MedMCQA\cite{pal2022medmcqa} and MedQA\cite{jin2021disease} focus on general medical knowledge, providing open-domain questions and multiple-choice answers derived from medical licensing and entrance exams. GeneTuring targets genomics-specific tasks, such as gene name conversion and nucleotide sequence alignment. Meanwhile, BioASQ\cite{tsatsaronis2015overview, nentidis2024overview} and PubMedQA~\cite{jin2019pubmedqa} incorporate supporting materials, such as PubMed articles, to answer domain-specific questions with formats ranging from yes/no to multi-class classifications. These datasets are crucial for benchmarking QA systems, as they provide domain-specific contexts and evaluation metrics that drive the development of more accurate and reliable models tailored to biomedical needs.

\textbf{Text Summarization (TS)} in the biomedical and healthcare is a critical application of natural language processing, enabling the condensation of complex medical texts into concise, informative summaries without compromising essential details. This task is particularly valuable in areas such as the summarization of the literature, the summarization of radiology reports, and the summarization of clinical notes. Among these, the summarization of radiology reports plays an essential role in transforming detailed imaging reports - including X-rays, CT scans, MRI scans, and ultrasounds - into easily understandable summaries. Datasets like MIMIC-CXR~\cite{johnson2019mimic} are instrumental in advancing this field, providing a large-scale resource with 473,057 chest X-ray images and 206,563 corresponding reports. Such data sets are essential for training and evaluating summarization models, offering domain-specific content and structured formats that drive improvements in both accuracy and reliability, ultimately enhancing clinical workflows and decision making.

\textbf{Information Extraction (IE)} in biomedicine involves organizing unstructured text into structured formats through tasks like named entity recognition (NER) and relation extraction (RE). Robust IE systems rely on high-quality datasets for training and evaluation. For instance: datasets such as BC5CDR~\cite{wei2016assessing}, NCBI-disease~\cite{dougan2014ncbi}, ChemProt~\cite{taboureau2010chemprot,kim2012chemprot,kringelum2016chemprot}, DDI~\cite{herrero2013ddi}, GAD~\cite{bravo2015extraction}, BC2GM~\cite{smith2008overview}, and JNLPBA~\cite{collier2004introduction} have become benchmarks for NER and RE tasks, addressing challenges involving diseases, chemicals, genes, and other biomedical entities. These datasets are essential benchmarks for tackling real-world biomedical challenges, enabling the development of more accurate and generalizable models.

LLMs have also shown potential in various biomedical tasks like coreference resolution and text classification. The effectiveness of these applications often depends on the availability of high-quality datasets. For coreference resolution, datasets such as MEDSTRACT~\cite{pustejovsky2002medstract}, FlySlip~\cite{gasperin2007annotation}, GENIA-MedCo~\cite{su2008coreference}, DrugNerAR~\cite{segura2010resolving}, BioNLP-ST’11 COREF~\cite{nguyen2011overview}, HANAPIN~\cite{batista2011building} and CRAFT-CR~\cite{cohen2017coreference} provide essential benchmarks for identifying links between mentions of the same entity in biomedical texts. Pre-trained models such as BioBERT~\cite{lee2020biobert} and SpanBERT~\cite{joshi2020spanbert} have achieved notable success in this domain. In text classification, datasets like HoC (comprising 1,580 manually annotated PubMed abstracts for multi-label classification of cancer hallmarks)~\cite{baker2016automatic} have been pivotal. 

In summary, the rapid progress in LLMs have transformed biomedical applications by improving data interpretation, knowledge extraction, and task automation. From question answering and text summarization to information extraction, LLMs have demonstrated their potential across a wide range of Bioinformatics-Specific tasks. Central to their success is the availability of high-quality, domain-specific datasets, which are indispensable for training, benchmarking, and refining these models to address real-world challenges. These datasets not only enhance the effectiveness of LLMs but also act as a driving force in advancing the field of bioinformatics and biomedicine. As the availability of diverse and richly annotated datasets continues to expand, they will fuel the integration of LLMs into increasingly complex and specialized applications. Looking to the future, combining Bioinformatics-Specific datasets with cutting-edge techniques promises to unlock groundbreaking solutions, enabling more precise, efficient, and scalable innovations that will shape the next generation of biomedical research and healthcare.

\subsection{Model Evolution and Key Milestones}
The evolution of LLMs in bioinformatics has marked a transformative journey. Initially developed for natural language processing tasks, these models, such as BERT~\cite{kenton2019bert} and GPT~\cite{radford2018improving}, have demonstrated remarkable potential in addressing challenges specific to the bioinformatics domain. Leveraging their ability to process and generate sequences, LLMs have been adapted for various biological data types, including DNA, RNA, proteins, and drug molecules~\cite{liu2024large}.

In genomics, models like DNABERT~\cite{ji2021dnabert} and GROVER~\cite{sanabria2023human} are trained on DNA sequences to predict functional regions, such as promoters and enhancers, and analyze mutations. Similarly, transcriptomics benefits from models like SpliceBERT~\cite{chen2023self} and RNA-FM~\cite{chen2022interpretable}, which assist in understanding RNA splicing and secondary structure prediction. For proteomics, PPLMs like ProtTrans\cite{elnaggar2007prottrans} and ProtGPT2~\cite{ferruz2022protgpt2} enhance predictions related to protein structure, function, and interactions. These advances are made possible by the foundational transformer architecture, which excels at processing sequential data. Fine-tuning these pre-trained models for domain-specific tasks extends their utility to applications drug discovery, where SMILES representations of molecules and protein sequences are integrated to predict interactions and properties.



A notable breakthrough in bioinformatics has been the AlphaFold series, which has applied cutting-edge machine learning to solve protein structure prediction challenges. AlphaFold2 (AF2) revolutionized structural biology with its unprecedented accuracy in predicting protein structures based solely on amino acid sequences. Its attention-based deep learning architecture captured intricate protein folding patterns, surpassing traditional physics-based and homology-modeling methods. By leveraging evolutionary information through multiple sequence alignments (MSAs), AF2 provided reliable predictions even in the absence of experimental data, significantly reducing the time and costs associated with obtaining protein structural information, accelerating advancements in drug discovery and functional genomics~\cite{bryant2022improved}.

Building on AF2’s success, AlphaFold3 (AF3) introduced groundbreaking capabilities, particularly in modeling protein complexes, including protein-peptide interactions. Transitioning from individual protein structure predictions to multi-component biological assemblies, AF3 addressed challenges protein-protein docking and protein-peptide interaction modeling. Through its template-based (TB) and template-free (TF) approaches, further extended the versatility and impact of the AlphaFold series~\cite{abramson2024accurate}. 

Key Features of AlphaFold3
Enhanced Accuracy in Complex Structures: AF3 excels in predicting protein-peptide complex structures, achieving a high percentage of accurate models in challenging scenarios; Innovative Template-Free Modeling: While maintaining strengths in template-based predictions, AF3 introduces powerful template-free algorithms that allow for diverse model generation with reliable accuracy, even in the absence of homologous structural data; Sophisticated Scoring and Ranking: AF3 integrates advanced scoring metrics such as DockQ and MolProbity, ensuring accurate evaluation of predicted structures. Its models show fewer issues like twisted peptides or cis non-proline residues, reflecting improved protein-like properties and geometric quality.

The progression from AF2 to AF3 reflects the iterative refinement of computational methods to address increasingly complex biological problems. While AF2 focused on individual protein structures, AF3 emphasizes dynamic interactions within biological systems, signaling a shift toward a more holistic understanding of molecular biology. These innovations underscore how machine learning continues to redefine bioinformatics, enabling accurate and efficient modeling of protein structures and interactions. The AlphaFold series exemplifies the potential for transformative breakthroughs in biology and medicine, paving the way for future applications in understanding complex biological systems.


\section{Applications in Bioinformatics Problems}

At the heart of LLMs lies the transformer architecture, which leverages an attention mechanism to manage word importance in context without the traditional constraints of recurrent (RNN) or convolutional (CNN) neural networks. The self-attention mechanism of transformers not only allows for robust parallelization and scalability but also excels at capturing long-range dependencies in text. In bioinformatics, the growing availability of extensive datasets across diverse tissues, species, and modalities presents both an opportunity and a challenge. Bioinformatics analysis typically seeks to uncover hidden relationships within vast amounts of data, which can be broadly categorized into two formats: molecular and cellular. Molecular data often consist of sequences—strings of four bases for DNA and RNA, and strings of twenty different amino acids for proteins. Cellular data, such as that from single-cell RNA-seq, single-cell ATAC-seq, or single-cell CITE-seq, typically takes the form of a count matrix with cells as rows and modalities as columns. While there are parallels between these data types and the structured data used in NLP, significant differences pose unique challenges for applying LLMs directly.  

A comprehensive LLM framework for bioinformatics involves three critical stages: data tokenization, model pre-training, and subsequent analyses. Due to the inherent differences between bioinformatics and conventional NLP data, researchers have been pioneering adaptations to the LLM architecture to better suit bioinformatics applications. The following section will provide a detailed overview of notable contributions in this evolving field.  

\subsection{Genome Level}

Genome data primarily provide molecular-level insights, focusing on the sequences of DNA and RNA. This format bears a strong resemblance to natural language, as it is structured as ordered sequences of strings. In this analogy, each nucleotide in a sequence read is akin to a character, each read is akin to a sentence, and the entire genome is comparable to the full article. To bridge the genome sequence and natural language, multiple studies try several ways to tokenize the genome sequence to make it similar to the concept of “word” in the natural language. To gain deeper insights into the functionalities of various genome segments, most studies apply the BERT (Bidirectional Encoder Representations from Transformers) as the core model, which excels in understanding the functions of a genome segment in relation to its surrounding genome region and is easily extended to different specific tasks by fine-tuning the model with specific dataset.  

\subsubsection{LLM for DNA Analysis}

In DNA analyses, biological sequences are encoded into structured tokens to facilitate effective model processing. A commonly adopted method involves tokenizing sequences into k-mers, typically ranging from 3 to 6 bases in length. This approach creates a vocabulary of k-mer permutations analogous to words in natural language, allowing the pre-trained model to decipher patterns within these k-mers. The choice of k directly affects the complexity and size of the resulting library, presenting a trade-off between modeling efficiency and accuracy. 

One of the pioneering methods, DNABERT \cite{ji2021dnabert}, tokenizes DNA sequence data using overlapping fixed-length k-mers, as well as the recently developed Nucleotide Transformer \cite{dalla2024nucleotide}. To enhance model efficiency, subsequent versions like DNABERT-2 \cite{zhou2023dnabert} and GROVER \cite{sanabria2023human} have employed Byte Pair Encoding (BPE) \cite{sennrich2015neural}, a statistical compression technique that iteratively merges the most frequently co-occurring genome segments. This method extends beyond fixed k-mer lengths, significantly improving the efficiency and generalizability of the models. HyenaDNA \cite{nguyen2024hyenadna} uses one-mer to tokenize the DNA sequence since it uses Hyena \cite{poli2023hyena} as the core model, which allows much longer input than BERT. Additionally, some models integrate supplementary data into their tokenization process; for instance, DNAGPT \cite{zhang2023dnagpt} incorporates species information, and MuLan-Methyl \cite{zeng2023mulan} combines sequence and taxonomy data into a natural language-like sentence to fully leverage existing LLM capabilities. 

In terms of pre-training approaches, many models utilize the BERT architecture with a masked learning method (MLM) for self-supervised training. To boost training efficiency, DNABERT incorporates the AdamW optimizer with fixed weight decay and applies dropout to the output layer. DNABERT-2 introduces enhancements such as Attention with Linear Biases (ALiBi) \cite{press2021train} and Flash Attention \cite{dao2022flashattention}. In contrast, the MuLan-Methyl framework integrates five fine-tuned language models (BERT and four variants) for the joint identification of DNA methylation sites, maintaining consistency with their original pre-training setups. DNABERT-S \cite{dnaberts2024} develops a contrastive learning-based method to help effectively cluster and separate different species. Some methods adopt other LLM models. For example, DNAGPT uses a GPT-based model and the next-token prediction for its pre-training, enabling it to forecast subsequent tokens based on previous ones. HyenaDNA uses Hyena, a new LLM model that allows a longer context input, to study long-range genomic sequence properties.

When applying these models to specific bioinformatics tasks, most integrate additional task-relevant data for fine-tuning. For instance, DNABERT and its derivatives utilize the Eukaryotic Promoter Database (EPDnew) \cite{dreos2013epd} to predict gene promoters, the ENCODE database \cite{encode2012integrated} for transcription factor binding site identification, and dbSNP for functional variant detection. MuLan-Methyl uses data from three main types of DNA methylation across multiple genomes for accurate predictions. Nucleotide Transformer includes multiple downstream tasks by fine-tuning the model with different datasets, like using histone ChIP-seq data \cite{encode2012integrated} for epigenetic marks prediction, using human enhancer elements data \cite{encode2020expanded} for enhancer sequence prediction, and using human annotated splice sites data \cite{harrow2012gencode} for splice site prediction. DNAGPT leverages data on polyadenylation signals and translation initiation sites for genomic signal and region recognition. Moreover, due to the generative nature of GPT, DNAGPT can also generate artificial human genomes without additional fine-tuning data. Without further fine-tuning, some methods use the embedding from the model directly. DNABERT-S can be used for species clustering and classification.

\subsubsection{LLM for RNA Analysis}

Unlike DNA, RNA analysis encompasses more complex and varied tasks, requiring tailored preprocessing strategies. RNABERT \cite{kalickirnabert}, mirroring the structure of DNABERT, employs the k-mer method for tokenizing RNA sequences. Given the typically shorter sequences of RNA compared to DNA, other models like SpliceBERT \cite{chen2024self}, RNA-MSM \cite{zhang2024multiple}, and RNA-FM \cite{chen2022interpretable} utilize single nucleotides (one-mers) for tokenization. In addition to sequence tokenization, these models often incorporate metadata during preprocessing. For instance, RNA-RBP \cite{yamada2022prediction} labels each sequence as positive or negative based on the presence of an RNA-binding protein (RBP) region, while SpliceBERT similarly labels sequences for RNA-splicing sites. RNA-MSM enhances its input by including multiple sequence alignments (MSA) \cite{wright2020rnacontest} to preserve the evolutionary history of sequences. 

The pre-training approach for RNA largely follows that of DNA, utilizing BERT's architecture and masked language modeling (MLM) for training. Specifically, RNA-MSM adopts a structure akin to AlphaFold2 \cite{jumper2021highly}, leveraging an MSA-transformer architecture. Depending on the target application, models are pre-trained with different datasets: RNABERT and RNA-MSM use sequences from the Rfam database, RNA-FM utilizes non-coding RNA sequences from RNAcentral \cite{rnacentral2021rnacentral}, and SpliceBERT is pre-trained with RNA sequences from 72 vertebrates available on the UCSC Genome Browser \cite{haeussler2019ucsc}. BERT-RBP is trained using the eCLIP-seq dataset, which includes RBP information \cite{pan2020rbpsuite}. 

Once trained, the BERT-based models process tokenized sequences to produce embeddings for each token. These embeddings are directly utilized in several applications; RNABERT employs them to classify RNAs from different families, while BERT-RBP uses them to predict RBP-binding sites. Furthermore, the attention maps generated as part of the model output play a critical role: SpliceBERT uses these maps to assess the impact of genetic variants on RNA splicing, BERT-RBP to analyze transcript region types and predict secondary structures, and RNA-MSM for secondary structure and solvent accessibility predictions. 

For task-specific enhancements, some models undergo fine-tuning with additional datasets. SpliceBERT, for example, is fine-tuned using a human Branchpoints dataset \cite{zhang2017bpp} to predict BP sites and the Spliceator dataset \cite{scalzitti2021spliceator} to assess splice sites across species. RNA-FM is fine-tuned with the PDB dataset \cite{singh2019rna} to facilitate RNA 3D structure reconstruction. 

\subsection{Gene Products Level}

With advances in single-cell technologies, researchers have gained enhanced insights into the functional roles and regulatory mechanisms of gene products within individual cells \cite{tanay2017scaling}. Single-cell RNA sequencing (scRNA-seq) data, which records the expression levels of various genes across individual cells, is particularly instrumental. Typically presented in a count matrix format, scRNA-seq data contrasts with sequence data; it lacks a natural order and contains numerical values rather than sequences of strings. Researchers have explored various methods to adapt this data for compatibility with LLMs, adjusting the representation of scRNA-seq data to harness the power of LLM methodologies.  

To adapt scRNA-seq data for LLM compatibility, researchers have devised various strategies. Models like Cell2Sentence \cite{levine2023cell2sentence}, tGPT \cite{shen2023generative}, and Geneformer \cite{theodoris2023transfer} employ a ranked sequence of gene symbols by expression level as inputs. ScGPT \cite{cui2024scgpt} and scBERT \cite{yang2022scbert} discretize gene expressions and treat them as tokens. Additionally, scGPT incorporates metadata for position embedding, while scBERT leverages gene2vec \cite{du2019gene2vec} to capture semantic similarities based on general co-expression. 

Some methods utilize a transformer-based architecture, which accommodates non-discrete inputs more flexibly. CIForm \cite{xu2023ciform} segments the gene expression vector of each cell into equal-length sub-vectors or patches. TOCICA \cite{chen2023transformer} groups gene expression into patches representing specific pathways, and ScTransSort \cite{jiao2023sctranssort} employs CNNs to generate gene-embedding patches, transforming the expression matrix into multiple 2D square patches. TransCluster \cite{song2022transcluster} uses linear discriminant analysis (LDA) to convert gene expression counts into embedding vectors. 

Unlike genome analyses, single-cell analyses adopt diverse model architectures for pre-training. For instance, Cell2Sentence, tGPT, and scGPT utilize GPT, whereas scBERT and Geneformer are based on BERT architecture. Transformer-based methods often integrate a linear classifier post-transformer and train a supervised model using cell types, as seen in CIForm, TOCICA, scTransSort, and TransCluster. 

The primary aim of scRNA LLM methodologies is to achieve accurate and generalized cell type annotations across various tissues and species. Supervised transformer-based methods use the pre-trained model directly for cell-type annotation. For instance, tGPT supports developmental lineage inference, and TOCICA enables interpretable dynamic trajectory analysis. LLM-based methods, post-pre-training, can be fine-tuned for specialized tasks or data-scarce scenarios. ScGPT is adaptable for tasks such as cell annotation, perturbation response prediction, batch effect correction, and gene regulatory network inference. Similarly, Geneformer can be fine-tuned to predict gene dosage sensitivity, chromatin dynamics, and gene network dynamics. 

\subsection{Epigenomics}

Decoding the information residing in the non-coding portion of the genome is one of the fundamental challenges in genomics \cite{preissl2023characterizing}. While substantial progress has been made in understanding the coding regions of the genome, non-coding regions remain poorly understood, particularly their roles in disrupting the regulatory syntax of DNA and their contributions to gene regulation. Existing LLMs, for example, Enformer \cite{avsec2021effective}, which take DNA sequences as input and perform downstream tasks, face two critical limitations: they cannot predict the functions of sequences in different cellular contexts, and they fail to incorporate 3D chromatin interaction data.


EpiGePT \cite{gao2024epigept} is a new LLM designed to overcome these challenges. It enables researchers to predict functionality in diverse cellular contexts and integrate 3D chromatin interaction data into genomic modeling. EpiGePT's architecture consists of four key components: a sequence module that analyzes DNA sequences, a transcription factor (TF) module that encodes cellular contexts, a transformer module that examines long-range interactions between DNA regions, and a prediction module that outputs context-specific gene regulation insights. To predict function in novel cellular contexts, EpiGePT employs its TF module, which represents the expression and binding activities of hundreds of transcription factors as a context-specific vector. This vector is then combined with DNA sequence features, which are tokenized into genomic bins—each representing a segment of the DNA sequence. These tokens, enriched with both sequence and context-specific TF features, form the input to the model, ensuring it captures both the local sequence information and the cellular context. This approach allows the model to treat each genomic bin as a token with embedded positional and biological context, leveraging the self-attention mechanism in the transformer module to learn long-range interactions and context-specific functionality. EpiGePT also addresses the challenge of incorporating 3D chromatin interaction data, which is critical for understanding long-range gene regulation. It guides the self-attention mechanism of its transformer module using ground truth 3D interaction data, such as HiChIP \cite{mumbach2016hichip} or Hi-C \cite{belton2012hi} loops. This alignment is achieved through a cosine similarity loss that adjusts the attention weights to reflect known 3D genomic interactions. By doing so, EpiGePT can model regulatory mechanisms, such as enhancer-promoter interactions, with higher fidelity than existing models.

\subsection{Protein Level}
Mass spectrometry (MS)-based proteomics focuses on characterizing proteins within complex biological samples~\cite{aebersold2003mass,shuken2023introduction}. Recent advancements in MS technology have enabled researchers to generate vast amounts of proteomics data~\cite{wang2024spdb}. However, the rapid growth in data volume presents significant analytical challenges. To tackle these issues, Ding et al. introduced PROTEUS, an LLM-based tool designed for automating proteomics data analysis and hypothesis generation~\cite{ding2024automating}. PROTEUS leverages a foundational LLM to integrate and coordinate existing bioinformatics tools, facilitating scientific discovery from raw proteomics data. Protein sequences share many similarities with natural language, and since breakthroughs have been achieved in applying NLP methods to protein sequence research, a variety of protein language models have emerged, differing in architecture, training strategies, and application scope~\cite{zhang2024scientific,sarumi2024large,xiao2024comprehensive}. Here, we outline the main types of protein language models and downstream tasks, each tailored to address distinct bioinformatics challenges in protein modeling, structure prediction, and functional annotation.

\subsubsection{Models for Protein LLM}
\textbf{Encoder-only models}, such as BERT-based models primarily designed for understanding protein sequences. These models excel in tasks that involve recognizing patterns within the sequences, making them suitable for protein classification, mutation effect prediction, and secondary structure analysis. Examples include  ESM 1b~\cite{rives2021biological}, ESM-1v~\cite{meier2021language},
ProteinBert~\cite{brandes2022proteinbert}, ProtTrans~\cite{elnaggar2021prottrans}, which leverage the bidirectional attention mechanisms of BERT to capture contextual relationships within amino acid sequences. 

\textbf{Decoder-only models}, similar to the GPT family in NLP, focus on generating new sequences based on learned distributions. In protein research, these models can be applied to generate synthetic protein sequences with desired properties or to design novel proteins. Models like
ProGen~\cite{madani2020progen},ProtGPT2~\cite{ferruz2022deep},
ZymCTRL~\cite{munsamy2022zymctrl},
 RITA~\cite{hesslow2022rita},
 IgLM~\cite{shuai2021generative}, 
 ProGen2~\cite{sternke2023proteinrl},
 and
 PoET~\cite{truong2023poet} are notable for their ability to produce diverse protein sequences that exhibit specific biochemical functions. This category is instrumental in protein engineering and synthetic biology, where the generation of novel, functional proteins is crucial~\cite{zhang2024scientific}.

\textbf{Encoder-decoder models} combine the strengths of both encoder-only and decoder-only architectures, making them highly adaptable to a range of protein-related tasks. They are particularly effective for sequence-to-sequence tasks, such as protein sequence alignment, where aligning amino acid sequences accurately is essential for understanding evolutionary relationships. These models can be fine-tuned for protein structure prediction or protein-protein interaction mapping, contributing to advancements in fields like drug discovery and disease diagnosis. The models include Fold2Seq~\cite{cao2021fold2seq},
 MSA2Prot~\cite{ram2022few},
Sgarbossaetal~\cite{sgarbossa2023generative},
 Leeetal~\cite{lee2023protein},
 LM-Design~\cite{zheng2023structure},
 MSA-Augmenter~\cite{zhang2023enhancing},
 ProstT5~\cite{heinzinger2023prostt5},
 xTrimoPGLM~\cite{chen2024xtrimopglm},
SS-pLM~\cite{serrano2023efficient},
 pAbT5~\cite{chu2023generative},
 ESM-GearNet-INR-MC~\cite{lee2023pre}.

\textbf{Multi-Modal Protein Models} integrate traditional protein language models with additional data types, such as structural and interaction information, to create powerful frameworks capable of analyzing both sequence and structural features simultaneously. By integrating textual protein sequences with structural annotations, these models enhance predictive capabilities for tasks such as 3D protein structure prediction, binding interaction analysis, and functional site identification. Frameworks like Multimodal Protein Representation Learning (MPRL)~\cite{duy2024multimodal} exemplify this approach by combining sequence information, 3D structural data, and functional annotations to capture the complex characteristics of proteins. For example, MPRL employs Evolutionary Scale Modeling (ESM-2)~\cite{lin2022language} for sequence analysis, Variational Graph Autoencoders (VGAE) for residue-level graphs, and PointNet Autoencoders (PAE) for 3D point cloud representations. This comprehensive data integration preserves both spatial and evolutionary aspects of proteins, allowing the model to generalize effectively across tasks like protein–ligand binding affinity prediction and protein fold classification. Similarly, Models like ProtTrans~\cite{elnaggar2021prottrans} and ESM~\cite{lin2022language} treat protein sequences as textual data, to learn rich embeddings that, when combined with 3D structural data, improve predictions of structure-function relationships. This multimodal synergy is essential for advancing protein engineering and drug discovery, mapping complex biological functions onto computational representations of proteins.

\subsubsection{Downstream Tasks for Protein LLM}
Protein modeling, especially through deep learning approaches, addresses a variety of critical tasks in biological research and medicine. For instance, deep learning methods are extensively applied in predicting protein-protein interactions (PPIs), which are fundamental for cellular functions~\cite{durham2023recent}. This prediction aids in understanding disease mechanisms, drug-target interactions, and the structural features of proteins that contribute to complex molecular pathways. The prediction of PPIs also enables the identification of novel therapeutic targets, providing significant insights for drug discovery and design. The typical models include AlphaFold~\cite{alquraishi2019alphafold}, AlphaFold 2~\cite{agarwal2024power}, AlphaFold 3~\cite{abramson2024accurate}, Graph-BERT~\cite{jha2023graph}, MARPPI~\cite{li2023marppi}.

Large-scale models also excel in predicting protein post-translational modifications (PTMs), which play essential roles in regulating protein function, stability, and cellular signaling~\cite{lee2023control}. Various machine learning models, including those based on transformers and neural networks, have been adapted to predict PTM sites with improved accuracy. For instance, the PTMGPT2 model~\cite{shrestha2024post}, developed by fine-tuning a GPT-2 architecture, leverages prompt-based approaches to identify subtle sequence motifs that correspond to PTM sites across diverse types~\cite{esmaili2023review}. By using custom tokens in its prompt, PTMGPT2 effectively captures sequence context and improves prediction accuracy, making it useful for identifying disease-associated mutations and potential drug targets.

Additionally, protein structure prediction remains a pivotal task in computational biology. It involves understanding how proteins fold and how their structures determine functions. Advanced models, like those using transformer architectures, facilitate the accurate prediction of protein structures, providing crucial information for synthetic biology, enzyme design, and therapeutic protein engineering~\cite{bertoline2023before}. These methods enable scientists to predict protein folding patterns and design novel proteins with specific functions, potentially revolutionizing fields like drug discovery and synthetic biology.The typical models include AlphaFold~\cite{alquraishi2019alphafold}, AlphaFold 2~\cite{agarwal2024power}, AlphaFold 3~\cite{abramson2024accurate}, ColabFold~\cite{kim2024easy}, Eigenfold~\cite{jing2023eigenfold}.

The development of protein large language models (Prot-LLMs) relies on diverse datasets that capture the complexity of protein sequences and functions. These datasets typically include unlabeled data for unsupervised pre-training, such as protein sequences from repositories like UniProt~\cite{uniprot2023uniprot}, AlphaFoldDB~\cite{nguyen2024data} which houses millions of protein sequences across species. For fine-tuning and evaluation, labeled datasets focus on specific protein characteristics, such as structure, function, and interactions. Examples include datasets for secondary structure prediction, protein-protein interaction networks, and specific post-translational modification (PTM) sites~\cite{lee2023dataset}. These labeled datasets enable Prot-LLMs to perform tasks like function annotation, PTM prediction, and protein structure modeling.

\subsection{Metabolomics}
Metabolomics represents the comprehensive analysis of the complete set of small-molecule metabolites within a biological system, providing a snapshot of the cellular biochemical status at a given time. This omics discipline is pivotal in elucidating the dynamic interactions between genotype and phenotype, as metabolites are the end-products of cellular processes and are directly involved in the regulation of biological functions. Metabolomics has emerged as a powerful tool in various areas of biological and medical research, including the identification of biomarkers for disease diagnosis, prognosis, and therapeutic monitoring~\cite{liu2024large}, as well as the elucidation of molecular mechanisms underlying disease pathogenesis. The integration of LLMs into metabolomics offers transformative potential for analyzing and interpreting metabolomic data. With their capacity to process vast amounts of textual and numerical information, LLMs, particularly transformer-based models adapted for biological data, have shown promise in metabolite identification and pathway analysis.


\subsubsection{Data Integration and Interpretation}

One of the most significant challenges in metabolomics is the integration and interpretation of large, complex datasets. LLMs can facilitate the integration of metabolomic data with other omics data (e.g., genomics, transcriptomics, proteomics) and clinical data, a challenge increasingly addressed by dynamic modeling approaches to enhance our understanding of metabolic phenotypes~\cite{lakrisenko2021dynamic}. By processing and analyzing these multi-omics datasets, LLMs can identify patterns and correlations that may not be apparent through traditional statistical methods. For instance, LLMs can be trained to predict the biological pathways and processes associated with specific metabolite profiles, thereby providing insights into the molecular mechanisms of disease.

Recent advances in multi-modal LLM architectures have addressed key challenges in data integration. The development of cross-attention mechanisms specifically designed for metabolomic data has improved the ability to handle heterogeneous data types. These mechanisms allow for simultaneous processing of spectral data, chemical structures, and biological annotations. However, significant challenges remain in handling the high dimensionality and sparsity of metabolomic data. Novel approaches incorporating dimensionality reduction techniques and attention-based feature selection have shown promise in managing these challenges while maintaining biological relevance.

\subsubsection{Biomarker Discovery and Validation}

The identification of robust biomarkers is a critical aspect of metabolomics, with applications in disease diagnosis, prognosis, and therapeutic monitoring. LLMs can be employed to analyze large datasets from clinical trials and cohort studies to identify potential biomarkers associated with specific disease states. Integrated deep learning frameworks have addressed challenges such as matching uncertainty and metabolite identification, enabling more reliable biomarker discovery and validation through the integration of diverse data sources~\cite{tian2023integrated}. This can lead to the development of more accurate and reliable biomarker panels for clinical use.

The validation of metabolomic biomarkers presents unique challenges that LLMs are increasingly equipped to address. Recent developments in uncertainty quantification for LLMs have improved the reliability of biomarker predictions. Statistical frameworks incorporating false discovery rate control and multiple hypothesis testing have been integrated into LLM-based biomarker discovery pipelines. Furthermore, the development of interpretable deep learning architectures has enhanced our ability to understand the biological mechanisms underlying identified biomarkers, leading to more robust validation processes.

\subsubsection{Metabolic Pathway Analysis and Drug Discovery}

Metabolomics data can provide valuable insights into the perturbations of metabolic pathways in disease states. LLMs exhibit remarkable capabilities in analyzing biological data, such as genomic sequences and protein structures, making them instrumental in identifying druggable targets and novel therapeutic compounds~\cite{tripathi2024large}. For example, LLMs can be trained to predict the effects of gene variants on enzyme activity and metabolic fluxes, thereby aiding in the identification of druggable targets. Additionally, LLMs can be used in the discovery of novel therapeutic compounds by predicting the binding affinity of small molecules to metabolic enzymes and pathways.

Advanced graph neural network architectures have emerged as powerful tools for metabolic pathway analysis when integrated with LLMs. These hybrid approaches can capture both the topological structure of metabolic networks and the chemical properties of individual metabolites. Recent developments in attention-based graph neural networks have improved our ability to predict metabolic flux distributions and identify regulatory bottlenecks. The integration of molecular docking simulations with LLM-based predictions has enhanced the accuracy of drug-target interaction predictions in metabolic pathways.

\subsubsection{Personalized Medicine}

The application of metabolomics in personalized medicine is rapidly gaining momentum, with the potential to tailor treatments to individual patients based on their metabolic profiles. LLMs can play a crucial role in this context by analyzing patient-specific metabolomic data in conjunction with genomic, proteomic, and clinical data to develop personalized treatment plans. For instance, LLMs can be used to predict the response of individual patients to specific therapies based on their metabolic profiles, thereby enabling the selection of the most effective treatment options.

\subsubsection{Literature Mining and Knowledge Discovery}

The vast amount of published literature in the field of metabolomics presents both an opportunity and a challenge for researchers. LLMs can be employed to mine this literature for relevant information, such as the identification of novel metabolites, the characterization of metabolic pathways, and the discovery of new biomarkers, addressing the challenge of synthesizing metabolomics research~\cite{kaddour2023challenges}. By processing and analyzing textual data from scientific articles, LLMs can generate hypotheses and identify trends that may guide future research directions.

\subsubsection{Quality Control and Data Standardization}

The reproducibility and comparability of metabolomics data are critical for the advancement of the field. Tools like the LargeMetabo package facilitate the reproducibility and standardization of large-scale metabolomics datasets, ensuring consistency across studies~\cite{bbac455}. LLMs can be used to standardize metabolomics data by identifying and correcting inconsistencies in data annotation, nomenclature, and reporting. Additionally, LLMs can assist in the development of quality control metrics and standards for metabolomics experiments, thereby improving the reliability and comparability of metabolomics data across different studies and platforms.

\subsubsection{Predictive Modeling and Simulation}

LLMs can be integrated with machine learning models to develop predictive models of metabolic pathways and networks. Advanced multivariate models, including machine learning techniques, have shown efficacy in analyzing metabolomics data to uncover predictive patterns of metabolic pathways~\cite{Vu2019Evaluation}. These models can be used to simulate the effects of genetic, environmental, and pharmacological perturbations on metabolic processes, thereby providing insights into the molecular mechanisms of disease and the potential outcomes of therapeutic interventions. Furthermore, LLMs can be used to predict the outcomes of metabolic engineering strategies in synthetic biology applications, such as the optimization of metabolic pathways for the production of biofuels, pharmaceuticals, and other valuable chemicals.

The integration of LLMs into metabolomics represents a significant advancement in the field, with the potential to enhance data analysis, interpretation, and knowledge discovery. By leveraging the power of LLMs, researchers can unlock the full potential of metabolomics data, leading to new insights into disease mechanisms, the development of novel therapeutic strategies, and the advancement of personalized medicine. As LLMs continue to evolve, their applications in metabolomics are expected to expand, further accelerating the pace of discovery and innovation in this exciting field.

\section{ Disease-Specific Bio-medical Applications}
The application of LLM technology to medical-related bioinformatics data offers significant potential to enhance various downstream biomedical tasks (Figure ~\ref{fig:main}(c)).

\subsection{Brain Aging and Brain Disease}

Large Language Models  are transforming the study and management of brain diseases by enabling innovative approaches to diagnosis, treatment, and knowledge discovery. These models excel in processing diverse data types including clinical notes, imaging studies, biological sequences, and brain signals, unlocking new possibilities for identifying disease patterns, predicting progression, and personalizing care. This section highlights the diverse applications of LLMs in brain diseases, focusing on three critical areas: clinical diagnostic support, therapeutic assistance, and information driven decision-making. Through these contributions, LLMs address longstanding challenges in managing complex neurological conditions, offering scalable and non-invasive solutions that enhance both research and clinical practice.

\subsubsection{Clinical Diagnosis Support}

Accurate and timely diagnosis is the foundation of effective medical care, particularly in complex and progressive conditions such as neurodegenerative diseases. The emergence of LLMs in healthcare offers transformative potential in clinical diagnostics by leveraging their advanced capabilities in processing diverse forms of unstructured data. From textual data to biological sequences and brain signals, LLMs excel at identifying patterns, extracting clinically relevant information, and supporting decision-making. Additionally, their ability to integrate multimodal data
has shown promise in improving diagnostic accuracy. This section explores how LLMs are applied to various data types crossing different brain diseases, highlighting their unique advantages and current challenges in clinical diagnosis.

\textbf{Textual data---Biomedical text} LLMs are increasingly applied to the analysis of biomedical textual data, including literature and electronic health records (EHRs). This form of biomedical textual data closely mirrors the fundamental structure of large language models. LLMs can identify significant insights within medical reports, enhancing diagnostic accuracy. In brain disease research, LLMs have been leveraged to diagnose conditions like seizures, Alzheimer’s disease (AD), headaches, strokes, Parkinson’s disease, and other neurodegenerative disorders using textual data from clinical notes, MRI reports, and neuropathological records.
For AD, LLMs provide a non-invasive, cost-effective, and scalable solution by analyzing unstructured data within EHRs. For example, Mao et al. demonstrated that the LLM can accurately predict MCI to AD progression using clinical notes as the early detection \cite{mao2023ad}. Feng et al. utilized LLMs to embed textual data in alignment with imaging data, significantly enhancing AD diagnosis through a multimodal approach \cite{feng2023large}.
Beyond AD, LLMs have also shown promise in managing epilepsy, with studies successfully classifying seizure-free patients and extracting seizure frequency and other critical information from clinical notes \cite{xie2022extracting}. Additionally, in a study analyzing neurodegenerative disorders at the Mayo Clinic, diagnostic accuracies of 76\%, 84\%, and 76\% were achieved using ChatGPT‐3.5, ChatGPT‐4, and Google Bard, respectively, underscoring the potential of LLMs in generating differential diagnoses for complex neuropathological cases \cite{koga2024evaluating}.
EHRs also include detailed MRI reports, which are critical in neurological diagnoses. Bastien Le Guellec et al. evaluated the performance of LLMs in extracting information from real-world emergency MRI reports, demonstrating high accuracy without requiring additional training \cite{le2024performance}. Similarly, Kanazawa et al. showed that a fine-tuned LLM could classify MRI reports such as no brain tumor, post-treatment brain tumor, and pre-treatment brain tumor with accuracy comparable to human readers \cite{kanzawa2024automated}. These results highlight the growing importance of LLMs in processing MRI reports, which are essential components of EHRs, further enhancing their utility in brain disease diagnosis and management.

\textbf{Textual data---Transcription text} In addition to text-based data, transcriptions from speech data are increasingly valuable for diagnosing brain diseases that impair linguistic abilities. Patients with AD, for example, often exhibit distinct speech patterns when describing images, including word-finding difficulties, grammatical errors, repetitive language, and incoherent narratives. The ADReSS Challenge dataset inspired the research community to develop automated methods to analyze speech, acoustic, and linguistic patterns in individuals to detect cognitive changes, frequently used in such studies \cite{valsaraj2021alzheimer, vats2021acoustic, bang2024alzheimer, agbavorpredicting}. LLMs outperform traditional methods like SVM, and Random Forest in this context. The existing work also shows that the combination of acoustic features with linguistic features for a multi-model can improve the performance, The maximum accuracy obtained by the acoustic feature is 64.5\%, and the BERT Model provides a classification accuracy of 79.1\% over the test dataset, the fusion of the acoustic model with the BERT Model shows an improvement of 6.1\% classification accuracy over the BERT Model \cite{vats2021acoustic}.
Linguistic analysis is also pivotal in diagnosing aphasia, a disorder commonly caused by left-hemisphere strokes. Chong et al. evaluated the clinical efficacy of LLM surprisals in a study where post-stroke aphasia patients narrated the story of Cinderella after reviewing a wordless picture book. The approach revealed significant potential for quantifying deficits and improving aphasia discourse assessment \cite{cong2024clinical}.

\textbf{Textual data---Text generation} In addition to biomedical text and speech data, recent advancements in text generation have further showcased the potential of large language models in clinical applications. Studies indicate that LLM-generated summaries are often preferred over those produced by human experts across various domains, including radiology reports, patient inquiries, progress notes, and doctor-patient dialogues \cite{van2024adapted}. This demonstrates the capacity of LLMs to synthesize complex clinical information effectively. Techniques such as Chain-of-Thought (CoT) prompting and text classification have been introduced to improve the confidence and precision of LLM outputs.
For example, when applied to neurologic cases, GPT-4 has shown promising results. By analyzing history and neurologic physical examination (H\&P) data from acute stroke cases, GPT-4 accurately localized lesions to specific brain regions and identified their size and number. This was achieved through Zero-Shot Chain-of-Thought and Text Classification prompting, highlighting the model's potential for advanced neuroanatomical reasoning \cite{lee2024gpt}. Similarly, in AD diagnostics, prompting LLMs with Clinical Chain-of-Thought frameworks has enabled them to generate detailed diagnostic rationales, demonstrating their ability to support reasoning-aware diagnostic frameworks \cite{kwon2024large}.

\textbf{Biological sequences} The process of DNA transcription to RNA, followed by translation into proteins, is fundamental to life and is often referred to as the Central Dogma of molecular biology. Many brain diseases, including AD, Parkinson’s disease (PD), autism spectrum disorder (ASD), and frontotemporal dementia (FTD), are closely associated with abnormalities in DNA, RNA, or protein sequences. To investigate the genetic and molecular mechanisms underlying these diseases, approaches such as Genome-Wide Association Studies (GWAS), transcriptome analysis, and proteomic profiling have been widely utilized. However, traditional methods often struggle to interpret the complex patterns present in these large-scale datasets. LLMs, with their advanced capabilities in processing sequential data, offer a transformative approach for analyzing biological sequences, enabling deeper insights into disease mechanisms and potential therapeutic targets. Several innovative LLMs have been developed for biological sequences. For DNA, models like Enformer \cite{avsec2021effective}, Nucleotide Transformer \cite{dalla2024nucleotide}, and DNABERT \cite{ji2021dnabert} have shown significant promise. For RNA, RNABERT \cite{akiyama2022informative}, RNAFM \cite{chen2022interpretable}, and RNA-MSM \cite{chen2022interpretable} focus on structural inference and functional predictions. For proteins, models like ProteinBERT \cite{brandes2022proteinbert}, ESM-1b \cite{rives2021biological}, and ProtST \cite{xu2023protst} have demonstrated capabilities in understanding sequence-function relationships. Despite these advances, the application of LLMs to reveal relationships between abnormalities in biological sequences and specific brain diseases remains limited. Notable exceptions include epiBrainLLM, proposed by Liu et al., which extracts genomic features from personal DNA sequences using a retained LLM framework and combines these features to enhance diagnosis \cite{liu2024leveraging}. This approach provides valuable insights into the causal pathways linking genotypes to brain measures and AD-related phenotypes. Another study utilized LLMs to predict protein phase transitions (PPTs) such as amyloid aggregation, a key pathological feature of age-related diseases like AD, demonstrating the potential of LLMs in advancing molecular-level understanding of neurodegenerative disorders \cite{frank2024leveraging}.

\textbf{Brain signal} Brain signal data, including sMRI, fMRI, and EEG, is critical for diagnosing and understanding various brain diseases. Abnormalities in these signals are key diagnostic indicators for conditions such as epilepsy, ADHD, and mental health disorders.
For epilepsy, EEG abnormalities such as seizures, spikes, and slowing patterns are widely used for diagnosis. A fine-tuned LLM, named EEG-GPT, was developed for classifying EEG signals as normal or abnormal, showing strong performance in identifying these patterns \cite{kim2024eeg}. Similarly, Liu et al. leveraged LLMs to guide affinity learning for rs-fMRI, enabling comprehensive brain function representation and improved diagnostic accuracy for brain diseases \cite{liu2024affinity}.
All of the LLM models above are based on the transformer architecture. Due to the long-range dependencies and temporal resolution in brain signals,  Mamba-based LLM also show its potential in this field. Behrouz and Hashemi proposed BrainMamba, an efficient encoder for modeling spatio-temporal dependencies in multivariate brain signals. It combines a time-series encoder for brain signals and a graph encoder for spatial relationships, making it versatile for neuroimaging data. With a selective state space model design, BrainMamba achieves linear time complexity, enabling training on large-scale datasets. Evaluations on seven real datasets across three modalities (fMRI, MEG, EEG) and tasks such as seizure, ADHD, and mental state detection show that BrainMamba outperforms baselines with lower time and memory requirements \cite{behrouz2024brain}.

\subsubsection{Therapeutic Assistance}

LLMs have demonstrated a strong capability to engage in conversations on daily life topics, personal matters, and specific concerns. When fine-tuned to provide empathetic and understanding responses, they hold significant potential as tools for companionship and emotional support. This capability is particularly valuable for individuals with dementia (PwD), who often experience social isolation. Research indicates that social isolation is strongly linked to an increased risk of developing dementia later in life \cite{holwerda2014feelings}. Addressing social isolation plays a vital role in mitigating cognitive decline among the elderly. Recent studies have explored the potential of LLMs to alleviate social isolation and provide therapeutic support. For example, Qi demonstrated that ChatGPT effectively reduces feelings of loneliness among older adults with mild cognitive impairment (MCI) by offering conversational engagement and cognitive stimulation \cite{qi2023chatgpt}. Similarly, Raile highlighted the dual role of ChatGPT as a complement to psychotherapy and an accessible entry point for individuals with mental health concerns who have yet to seek professional help \cite{raile2024usefulness}. These findings suggest that LLMs can serve as valuable tools to support mental health and cognitive functioning in vulnerable populations.

In the context of neurodegenerative diseases, wearable devices integrated with AI technologies offer promising avenues for continuous monitoring and personalized care. Mohammed and Venkataraman introduced an AI-powered wearable device that leveraging LLMs to monitor the daily activities of patients with PD by analyzing multimodal data such as tremors, movements, and posture \cite{mohammed2023innovative}. This approach enables real-time and personalized assessments of disease progression, potentially enhancing patient care and quality of life.

Language impairments, such as aphasia, present significant challenges in communication. Manir et al. utilized BERT models to predict and complete sentences for individuals with aphasia, thereby improving the accuracy of speech prediction \cite{manir2024llm}. This approach benefits caregivers and speech therapists by facilitating more effective communication strategies and supporting rehabilitation efforts.

Brain-computer interfaces (BCIs) further exemplify the integration of advanced AI techniques into healthcare. Over recent decades, BCIs have provided novel solutions for various neurodegenerative disorders, including AD \cite{liberati2012toward} and PD \cite{miladinovic2020evaluation}. The incorporation of advanced AI algorithms, such as machine learning and deep learning, has significantly enhanced BCI performance, improving neuroergonomic systems, human-robot interactions, and robotic-assisted surgeries \cite{li2016human, cao2020review}. Notably, integrating LLMs with BCIs introduces unique opportunities, such as reliably comprehending users’ emotional states to create emotionally aware conversational agents \cite{sorino2024ariel} and decoding attempted speech from the brain activity of paralyzed patients \cite{jimenez2024arithmetic}. These advancements highlight the transformative potential of LLMs in facilitating communication and enhancing the quality of life for individuals with severe disabilities.

Collectively, these studies underscore the versatile applications of LLMs as therapeutic assistance in brain diseases. By enhancing social interaction, providing cognitive support, enabling continuous monitoring, and assisting in communication, LLMs represent a promising avenue for improving patient outcomes and overall quality of life.

\subsubsection{Information Driven
Decision-making}

LLMs have proven to be valuable tools for information retrieval, serving as vast repositories of knowledge. Saeidnia et al. reported that dementia caregivers expressed positive feedback on ChatGPT's responses to non-clinical questions related to the daily lives of individuals with dementia \cite{saeidnia2024evaluation}. This suggests that LLMs can support caregivers by providing accessible and practical information to manage everyday challenges. However, concerns remain about the depth and accuracy of medical information provided by LLMs. Studies comparing ChatGPT with traditional search engines have found limitations in the quality of responses, describing them as accurate but lacking in comprehensiveness \cite{hristidis2023chatgpt}. These findings suggest that while LLMs can address basic queries, their applicability in complex medical contexts requires further refinement.
One solution to these limitations is fine-tuning LLMs using domain-specific data. For example, models trained on medical journals and textbooks have demonstrated improved performance in handling specialized medical queries \cite{wu2024pmc}. In Alzheimer’s research, GPT-4-based tools have been developed to autonomously collect, process, and analyze health information, illustrating how customization can enhance the relevance and precision of information retrieval in specific medical domains \cite{dai2023ad}.

\subsection{Cancer Treated by Radiation Therapy}

LLMs have emerged as powerful tools in cancer research, offering innovative solutions for diagnosis, treatment planning, and biological insights. By processing vast datasets of scientific literature, clinical trial results, and genomic information, LLMs can facilitate the identification of novel biomarkers and treatment strategies. LLM-driven multimodal approaches have also enhanced target volume contouring in radiation oncology, integrating imaging data with clinical notes for improved precision\cite{oh2024llm, oh2024mixture}. In radiobiology, these models contribute to understanding the complex interplay between radiation and cellular processes, informing the development of personalized treatment regimens\cite{wang2024fine}. Recent studies also explore the application of LLMs across chemotherapy, surgery, radiotherapy, and immunotherapy, demonstrating their versatility and potential in advancing oncology research.

Multimodal large language models (MLLMs) that integrate imaging analysis with natural language processing have shown promising results in automated organ-at-risk (OAR) and target volume delineation, achieving expert-level performance \cite{WOS:001342098500028}. These models can process multiple imaging modalities—CT, MRI, and PET—simultaneously while incorporating clinical notes and radiology reports to improve contour accuracy. Additionally, LLMs are being utilized for dose prediction \cite{WOS:001319329400001}, where they have the potential to suggest optimal dose distributions for patients. Recent studies have explored their application in adaptive radiotherapy, where LLMs show potential in processing daily imaging data to recommend plan adaptations based on anatomical changes. Integrating LLMs with knowledge-based planning systems has also enhanced the quality of treatment plans by leveraging insights from large databases of previously treated cases. Furthermore, LLMs demonstrate potential in predicting treatment outcomes and toxicity risks by analyzing patient-specific factors, enabling more personalized treatment approaches.

In clinical practice, LLMs are proving useful in automating routine tasks and supporting complex decision-making\cite{HOLMES2024e515}. Tools like ChatGPT have been piloted for generating comprehensive patient case reports, improving the efficiency of clinical documentation. Furthermore, LLMs have shown promise in extracting discrete data elements from clinical notes, aiding in the creation of robust cancer databases. They were evaluated for supporting personalized oncology by recommending clinical trials for head and neck cancer and offering decision support for treatment planning. However, these applications require rigorous validation to ensure the accuracy and reliability of their outputs.

In education, LLMs are transforming how knowledge is disseminated and acquired in oncology. Educational chatbots tailored to radiation oncology can simulate patient interactions, helping trainees refine their communication skills\cite{hao2024retrospective}. Additionally, LLMs assist in evaluating radiotherapy plans and providing structured feedback, as demonstrated by recent studies. These models foster a more interactive and adaptive learning environment, enabling personalized educational experiences for medical physicists, oncologists, and other healthcare professionals\cite{wang2025recentevaluationperformancellms}. Despite challenges such as ensuring content accuracy and avoiding the propagation of biases, the integration of LLMs into educational frameworks holds the potential to enhance competency and foster innovation in cancer care.


\subsection{Infectious Diseases}

\subsubsection{Disease Prediction and Vaccine Efficacy Analysis}

LLMs such as GPT-3 and GPT-4, have emerged as powerful tools in disease prediction and vaccine efficacy analysis. By processing vast datasets, including biomedical records and epidemiological trends, LLMs can model the spread of infectious diseases, predict vaccination outcomes, and assist in assessing vaccine effectiveness. For example, neural networks combined with logistic regression have been applied to predict influenza vaccination outcomes, achieving significant accuracy based on demographic and clinical data \cite{trtica-majnaric2010prediction}. In the context of pediatric respiratory diseases, ChatGPT has been used to generate insights and recommendations for reducing severe cases post-COVID-19, highlighting the adaptability of LLMs in addressing real-world healthcare issues \cite{alhasan2023mitigating}. Additionally, machine learning algorithms based on clinical features have been validated for predicting influenza infection in patients with influenza-like illness (ILI), illustrating the role of LLMs in early diagnosis and targeted intervention \cite{hung2023developing}. LLMs are also instrumental in identifying immune biomarkers that predict vaccine responsiveness, as seen in studies exploring apoptosis and other immune markers to assess influenza vaccine efficacy \cite{furman2013apoptosis}. Furthermore, LLMs have been applied to the extraction and analysis of post-marketing adverse events from the Vaccine Adverse Event Reporting System (VAERS), providing valuable insights into vaccine safety and public health implications \cite{du2021extracting}. The use of machine learning for seasonal antigenic prediction, particularly for influenza A H3N2, demonstrates LLMs' potential in tracking viral evolution and optimizing vaccine design to address emerging strains \cite{shah2024seasonal}. As LLM technology continues to advance, its application in disease prediction and vaccine efficacy is expected to become increasingly essential in public health management and disease prevention strategies.

\subsubsection{Vaccine Adherence and Risk Prediction}

Machine learning and feature selection techniques, facilitated by LLMs, are essential in analyzing vaccine adherence patterns and identifying factors influencing vaccination rates. These methods allow researchers to process large, complex datasets, uncovering demographic and health-related variables that impact vaccine adherence and risk prediction. For example, machine learning models have been applied to assess low adherence to influenza vaccination among adults with cardiovascular disease, offering insights into the unique barriers to vaccination faced by high-risk groups \cite{kim2021machine}. Real-time data from online self-reports, such as social media posts, have also been used to track influenza vaccine uptake, providing valuable insights into public sentiment and adherence trends \cite{huang2019can}. Furthermore, sociodemographic predictors of vaccine acceptance, especially during the COVID-19 pandemic, have been studied extensively. For instance, machine learning has been used to explore the influence of variables like education level, income, and geographic location on vaccine hesitancy across various populations \cite{mondal2021sociodemographic}. In addition, validated scales such as the Parental Attitude about Childhood Vaccination Scale have been enhanced with feature selection techniques, refining our understanding of factors associated with vaccine acceptance and hesitancy \cite{ghazy2023external}. Other studies emphasize the broader implications of vaccine hesitancy by analyzing attitudes toward COVID-19 vaccinations across continents, highlighting the variability in hesitancy due to cultural and regional factors \cite{sammut2023covid19}. Lastly, comparative studies on flu vaccine uptake pre- and post-COVID-19 leverage machine learning to identify shifts in adherence patterns and factors that predict vaccination behavior over time \cite{skyles2024comparison}. Together, these advancements in machine learning and feature selection provide a comprehensive understanding of vaccine adherence, informing targeted public health strategies to improve vaccination rates.

\subsubsection{Biomarker Analysis and Antigen Prediction}

LLMs and machine learning approaches are increasingly being applied to analyze biomarkers and predict antigenic variations, which are essential for understanding immune responses and optimizing vaccine design. In biomarker analysis, studies have leveraged LLMs to investigate genetic relationships and autoimmune markers, helping to elucidate the factors that influence vaccination outcomes and susceptibility to infectious diseases \cite{mcgarvey2014silico}. For example, differential network centrality analysis and feature selection techniques have been employed to identify key susceptibility hubs within biological networks, offering insights into factors that contribute to immune response variability \cite{lareau2015differential}. 

Additionally, antigenic prediction plays a crucial role in designing effective influenza vaccines, especially for rapidly evolving strains. Statistical analyses of antigenic similarity, such as those conducted for influenza A (H3N2), highlight the potential of machine learning models in mapping antigenic drift and optimizing strain selection for seasonal vaccines \cite{adabor2021statistical}. Moreover, cellular correlates of protection identified through human influenza virus challenges have advanced our understanding of immune responses to oral vaccines, demonstrating the applicability of machine learning models in immune signature identification \cite{mcilwain2021human}. Blood inflammatory biomarkers have also been analyzed to differentiate COVID-19 from influenza cases, showcasing the predictive power of LLMs in clinical biomarker differentiation \cite{luciani2023blood}. Seasonal antigenic prediction, particularly for influenza A H3N2, has benefited from machine learning approaches that help forecast viral evolution, supporting timely vaccine updates \cite{shah2024seasonal}. Finally, phylogenetic analyses have identified optimal influenza virus candidates for seasonal vaccines, underscoring the significance of LLMs in guiding vaccine development against anticipated strains \cite{hayati2023phylogenetic}.

\subsubsection{Vaccine Recommendation and Immune Response}

LLMs are increasingly leveraged in vaccine recommendation and immune response studies, especially in analyzing antigenicity and optimizing vaccine strain selection. For instance, the MAIVeSS platform utilizes LLMs to streamline the selection of high-yield, antigenically matched viruses for seasonal influenza vaccines, a critical step in addressing annual viral mutations \cite{gao2024maivess}. In populations with specific health conditions, such as HIV, LLMs have been applied to predict the immunogenicity of trivalent inactivated influenza vaccines, revealing key biomarkers and immune signatures that inform personalized vaccination strategies \cite{cotugno2020artificial}. 

Antigenicity prediction models have also employed convolutional neural networks to optimize vaccine recommendations for influenza virus A (H3N2), facilitating the identification of effective vaccine strains through detailed computational modeling \cite{lee2020antigenicity}. Furthermore, temporal topic models generated from clinical text data allow for a more nuanced understanding of immune responses over time, especially in relation to patient health history and demographic factors, enhancing the precision of vaccine recommendations \cite{meaney2022comparison}. Finally, studies on COVID-19 vaccine hesitancy among populations already immunized for influenza underscore the relevance of LLMs in analyzing and addressing hesitancy factors, which is vital for improving adherence to vaccination programs \cite{valerio2022high}. Together, these applications illustrate the potential of LLM-based approaches in advancing vaccine recommendation processes and tailoring immune response strategies.

\subsubsection{Sentiment Analysis and Public Attitude Research on Social Media}

LLM techniques are widely used in sentiment analysis to assess public attitudes towards vaccines, particularly through social media data. This approach provides insights into public sentiment trends and identifies factors contributing to vaccine hesitancy or acceptance. For instance, social media analysis of public messaging around influenza vaccination from 2017 to 2023 has shown how sentiment fluctuates in response to vaccine news, policy changes, and health crises, offering a longitudinal view of public perception \cite{ng2023examininga}. Similarly, negative sentiments related to influenza vaccines, analyzed from over 260,000 Twitter posts, highlight recurring concerns and misconceptions that can be addressed through targeted public health messaging \cite{ng2023examining}. 

Beyond social media, predictive models using smartwatch and smartphone data can monitor side effects and public reactions post-vaccination, enhancing our understanding of vaccine safety perceptions \cite{levi2024prediction}. The FDA's Biologics Effectiveness and Safety Initiative also uses NLP to process unstructured data, identifying adverse events associated with vaccines and contributing to more accurate public health responses \cite{deady2021food}. Additionally, integrating immune cell population data and gene expression with CpG methylation patterns offers insights into immune responses that can correlate with public attitudes, informing data-driven interventions \cite{zimmermann2017integration}. These findings underscore the utility of LLMs in sentiment analysis, enabling public health authorities to monitor and respond to vaccine-related concerns effectively.

\subsubsection{Epidemiology and Public Health Data Analysis}

Machine learning and large datasets have profoundly impacted epidemiology and public health, enabling the analysis of disease patterns, risk factors, and vaccination responses. Studies integrating socioeconomic, health, and safety data have examined how these factors affect COVID-19 spread, offering insights into the influence of demographics like income and healthcare access on infection rates \cite{galvan2020can}. Projects like the Human Vaccines Project also leverage large datasets to map immune responses across populations, enhancing our understanding of vaccine design and immunology \cite{wooden2018human}.

The use of wearable sensors in epidemiological studies, as demonstrated in the WE SENSE protocol, facilitates early detection of viral infections by analyzing real-time health metrics, thus supporting timely public health interventions \cite{hadid2023we}. Pneumonia research, such as the work by the CAPNETZ study group, highlights unmet needs in understanding disease mechanisms, emphasizing the need for targeted data collection and analysis in developing effective treatment and intervention strategies \cite{pletz2022unmet}. Additionally, sociodemographic studies on COVID-19 vaccine acceptance reveal how age, gender, and education level impact vaccine uptake, providing crucial insights for public health policy \cite{mondal2021sociodemographic}. These applications underscore the essential role of data-driven approaches in epidemiology and public health to improve disease prevention and health policy.

\section{Drug Discovery and Development}

\subsection{Drug Target Identification}
Drug discovery is a resource-intensive and time-consuming process, often spanning 7 to 20 years from initial development to market approval \cite{{berdigaliyev_overview_2020},{cummings_alzheimers_2024}}. Central to this process is drug-target interaction (DTI) identification, which involves pinpointing molecules implicated in disease mechanisms. Traditional methods, including genomics, proteomics, RNAi, and molecular docking, have been instrumental but face limitations in cost, scalability, and adaptability to complex biological systems. 

Recent advancements in computational techniques, such as machine learning\cite{{sadybekov_computational_2023},{wang_predicting_2024},{frey_neural_2023}}, knowledge graph-based methods\cite{{huang_foundation_2024},{singhal_large_2023}}, and molecular docking simulations, driven by the rapid growth of large-scale biomedical datasets\cite{{zdrazil_chembl_2024},{koscielny_open_2017},wishart_drugbank_2018}, have significantly advanced DTI prediction. Beyond these methods, recent breakthroughs in LLMs and BioLMs represent a paradigm shift, enabling the integration and analysis of vast, heterogeneous datasets—including molecular data, biological networks, and scientific literature—while storing drug-related background knowledge through extensive pretraining \cite{{pun_ai-powered_2023},{savage_drug_2023},{sheikholeslami_druggen_2024},{m_bran_augmenting_2024},{boiko_autonomous_2023}}. This section provides an overview of LLM-based approaches for DTI prediction, categorized based on the type of data they utilize: sequence data, structural data, and relationship data, with the latter primarily derived from knowledge graphs.

Sequence data, including amino acid sequences for proteins and Simplified Molecular Input Line Entry System (SMILES) representations for drugs, plays a central role in single-modal methods for DTI prediction. Pretrained language models (PLMs), such as PharmBERT\cite{valizadehaslani_pharmbert_2023}, BioBERT\cite{lee_biobert_2020}, and ProteinBERT\cite{brandes_proteinbert_2022}, have been widely utilized to extract meaningful representations from such data, enabling efficient and accurate predictions. For instance, DTI-LM \cite{chaves_tx-llm_2024} addresses the cold-start problem by utilizing PLMs to predict DTIs based solely on molecular and protein sequences, enabling accurate predictions for novel drugs and uncharacterized targets. Similarly, ConPLex \cite{singh_contrastive_2023} generates co-embeddings of drugs and target proteins, achieving broad generalization to unseen proteins and over 10× faster inference compared to traditional sequence-based methods, making it ideal for tasks like drug repurposing and high-throughput screening. Yang et al. \cite{yang_advancing_2024} further enhance DTI prediction by introducing high-frequency amino acid subsequence embedding and transfer learning, capturing functional interaction units and shared features across large datasets. Additionally, TransDTI \cite{kalakoti_transdti_2022} employs transformer-based language models to classify drug-target interactions into active, inactive, or intermediate categories, offering competitive performance. Despite their advantages, single-modal methods are limited by their reliance on sequence data alone, making it challenging to capture interactions involving spatial, structural, or contextual dependencies.

To address the limitations of single-modal approaches, multimodal frameworks integrate diverse data types—such as molecular graphs, protein sequences, and structural data—offering a more comprehensive understanding of DTIs. DrugLAMP\cite{luo2024accurate} exemplifies this integration, utilizing Pocket-Guided Co-Attention (PGCA) and Paired Multi-Modal Attention (PMMA) to fuse molecular graphs with sequence data, achieving nuanced molecular interaction predictions. PGraphDTA \cite{bal_pgraphdta_2024} incorporates 3D contact maps alongside protein sequences, outperforming sequence-only methods when structural data is available. Beyond predictive accuracy, multimodal frameworks like CGPDTA\cite{fan_cgpdta_2025} enhance interpretability by integrating interaction networks, providing insights into biological mechanisms. DrugChat \cite{liang_multi-modal_2024} combines prompt-based learning with sequence data and textual inputs. Pretrained on three datasets, it predicts indications, mechanisms of action, and pharmacodynamics while dynamically generating textual outputs in response to user prompts. This eliminates the need for retraining and enables flexible, interactive exploration of drug mechanisms. Similarly, DrugReAlign\cite{ma_y-mol_2024} employs a multi-source prompting approach that integrates diverse and reliable data inputs to integrate textual and structural data, enhancing drug repurposing efforts.

Beyond structural data, KG-based models leverage semantic relationships, such as shared pathways, biological processes, and functional annotations, along with diverse data sources to achieve competitive performance in DTI predictions. Y-Mol \cite{ma_y-mol_2024} enhances biomedical reasoning by integrating multiscale biomedical knowledge and using LLaMA2 as its base LLM. It learns from publications, knowledge graphs, and synthetic data, enriched by three types of drug-oriented prompts: description-based, semantic-based, and template-based, enabling robust drug interaction analysis. 
Similarly, the multi-agent framework DrugAgent \cite{inoue_drugagent_2024} advances drug repurposing by combining AI-driven DTI models, knowledge graph extraction from databases (e.g., DrugBank, CTD\cite{davis_comparative_2021}), and literature-based validation. This framework integrates diverse data sources to streamline repurposing candidate identification, enhancing efficiency, interpretability, and cost-effectiveness. Together, these models boost predictive power while fostering collaboration and refinement.

\subsection{Molecular Docking and Drug Design}

The advanced reasoning capabilities of large language models  have enabled their application in biological and medical fields, demonstrating significant potential to accelerate drug discovery and screening processes~\cite{gong2023evaluating,zhang2024generalist}. Built upon the transformer architecture from Natural Language Processing (NLP), biology-focused language models have emerged as powerful tools to support both sequence-based and structure-based drug design~\cite{chen2023sequence,zhang2023protein,fang2023method}. By utilizing their strengths in text summarization and contextual understanding, these models can integrate information from diverse sources, such as scientific literatures, patent databases, and specialized datasets, to provide comprehensive analyses and insights into protein sequences, structures, binding pockets, and interaction sites~\cite{chakraborty2023artificial}. Moreover, protein language models and other transformer-based models are being applied to exploit unknown structural information in structure-based drug design (SBDD)~\cite{zhang2023protein,fang2023method}.

Molecular docking, a pivotal component of Structure-Based Drug Design (SBDD), necessitates three-dimensional protein structures and precise binding site information to calculate binding affinities during in silico virtual screening~\cite{shaker2021silico}. LLMs have shown potential to enhance various aspects of molecular docking, including docking input file generation, binding site prediction, and protein structure prediction~\cite{zhang2023protein,fang2023method,sharma2023chatgpt}. AutoDock is a widely adopted software for molecular docking~\cite{morris2008using}. For high-throughput drug screening, it is necessary to generate docking commands in text file format and execute them in the terminal. Sharma et al. demonstrated the capability of ChatGPT to generate AutoDock input files and basic molecular docking scripts~\cite{sharma2023chatgpt}. Another notable example is DrugChat, a ChatGPT-like LLM for drug molecule graphs developed by Liang et al. With the input of compound molecule graphs and appropriate prompts, DrugChat is able to generate insightful responses\cite{liang2023drugchat}.

Ligand binding site identification and prediction are essential for drug design. Due to the limited availability of experimentally determined protein crystal structures and incomplete protein structural knowledge, ligand binding site identification can be tough. Zhang et al. addressed this limitation through LaMPSite, an algorithm powered by EMS-2 protein language model, which only requires protein sequences and ligand molecular graphs as inputs without any protein structural information~\cite{zhang2023protein}. This approach achieved comparable performance to those methods requiring 3D protein structures in benchmark evaluations. Regarding of deficiency of reliable protein structure, protein language models have been applied for protein structure prediction as well. For example, Fang et al. introduced HelixFold-Single, a multiple-sequence-alignment-free protein structure predictor~\cite{fang2023method}. Unlike AlphaFold2, which enhances prediction accuracy by relying on multiple sequence alignments of homologous proteins, HelixFold-Single adopts a more efficient approach. It leverages large-scale protein language model training on the primary structures of proteins while integrating key components from AlphaFold2 for protein geometry.

Recent advancements in protein-ligand binding prediction methods have further enhanced screening efficiency and accuracy. Shen et al. developed RTMScore, which integrated Graph Transformer to extract structural features of protein and molecule, using 3D residue graphs of protein and 2D molecular graphs as inputs for protein-ligand binding pose prediction~\cite{shen2022boosting}. RTMScore outperformed many state-of-the-art docking software including Autodock Vina~\cite{trott2010autodock} , DeepBSP~\cite{bao2021deepbsp}, and DeepDock~\cite{mendez2021geometric} in performing virtual screening tasks. Another notable development is ConPlex, a sequence-based drug-target interaction (DTI) prediction method introduced by Singh et al~\cite{singh2023contrastive}. By employing representations generated from pre-trained protein language models (PLMs) as the inputs, ConPlex benefits from a larger corpus of single protein sequences and alleviates the problem of limited DTI training data. Additionally, contrastive learning was adopted to address the fine-grained issues by employing contrastive coembedding, which is able to co-locate the proteins and the targets in a shared latent space. Thus, a high specificity can be achieved by separating the true interacting patterns and decoys. According to contrastive training results, the effective size between true and decoy scores was largely increased.

Through automated data extraction and normalization, LLMs can greatly improve the efficiency and accuracy of drug property predictions. With ADMET (Absorption, Distribution, Metabolism, Excretion, and Toxicity) analysis, LLMs can also help distinguish the compounds possessing favorable profiles from those showing adverse characters and allow developing the most promising drug candidates during the pipeline process. For instance, PharmaBench achieves this through its multi-intelligence system, whose core function is to extract ADMET-related data from multiple public databases using LLMs~\cite{niu2024pharmabench}. Beyond ADMET analysis, LLMs like ChatGPT have expanded their capabilities to predict and analyze other features of drugs, including pharmacodynamics and pharmacokinetics, thus providing a comprehensive evaluation of potential drug candidates~\cite{chakraborty2023artificial}. LLMs powerfully accelerate the drug development pipeline by fastening data analysis, enhancing prediction accuracy, and offering all-rounded drug property evaluation, which in turn reduces both the time and resources needed for drug discovery and improves the chances of coming up with a successful drug candidate.

\section{Immunology and Vaccine Development}

Large Language Models , including GPT-based architectures, have transformed the field of immunology and vaccine development by enabling advanced analyses of large, complex datasets. These models, combined with machine learning, NLP, and feature selection techniques, facilitate the identification of immune biomarkers, prediction of vaccine efficacy, understanding of vaccine hesitancy, and real-time monitoring of adverse events. This review synthesizes recent research highlighting the critical role of LLMs in advancing vaccine science, with a focus on immune response analysis, vaccine development, efficacy prediction, safety, and public attitudes.

\subsection{Immune Response Analysis and Biomarker Research}

Analyzing immune responses and identifying biomarkers are critical for understanding the efficacy and mechanisms of vaccines. Large Language Models , integrated with advanced computational techniques, play a key role in processing and interpreting complex datasets to uncover immune signatures and their correlation with vaccination outcomes. For example, LLMs can efficiently analyze high-dimensional datasets, such as the FluPRINT dataset, which provides a multidimensional analysis of the immune system's imprint following influenza vaccination, revealing variability in immune responses across individuals \cite{tomic2019fluprint}. By leveraging LLMs, researchers can extract patterns and relationships from immune cell populations, mRNA sequencing, and CpG methylation data, leading to more accurate predictions of humoral immunity and highlighting the impact of gene expression and epigenetic modifications on vaccine-induced immunity \cite{zimmermann2017integration}.

Automated systems like SIMON utilize machine learning, augmented by LLMs for text-based data extraction and interpretation, to reveal immune signatures that predict vaccine responsiveness, providing deeper insights into immune mechanisms \cite{tomic2019simon}. Furthermore, LLMs facilitate the integration of multi-level models that incorporate gene expression interaction networks to predict antibody responses to vaccines, enabling the precise identification of immune predictors \cite{parvandeh2019multilevel}. For biomarker analysis, LLMs contribute to identifying apoptosis and inflammatory responses through their ability to process vast quantities of biological literature and experimental data, as seen in studies linking immune biomarkers with influenza vaccine responsiveness \cite{furman2013apoptosis}. They also assist in differentiating immune responses to COVID-19 and influenza infections by analyzing blood inflammatory biomarkers and clinical data at scale \cite{luciani2023blood}.

Additionally, human influenza virus challenge models, supported by LLM-driven analysis of experimental outcomes, have identified cellular correlates of protection, advancing our understanding of immune responses to oral vaccines \cite{mcilwain2021human}. LLMs streamline the analysis of complex immune response datasets, ensuring faster identification of key findings and improving collaboration across interdisciplinary research teams.

\subsection{Vaccine Development and Recommendation Models}

The development and optimization of vaccines rely on computational models to predict vaccine efficacy, identify suitable strains, and recommend antigenically matched candidates. Large Language Models  have become invaluable tools in this domain by enhancing the ability to process and analyze vast datasets, extract patterns from biomedical literature, and improve antigenic prediction models. Neural networks and logistic regression have traditionally been applied to predict influenza vaccination outcomes, providing robust frameworks for assessing vaccine effectiveness based on demographic and clinical data \cite{trtica-majnaric2010prediction}. With the integration of LLMs, these predictive models can be further refined by incorporating insights derived from textual datasets, such as clinical notes, trial reports, and patient feedback.

In silico approaches, combined with LLM-based text mining, enable the analysis of autoimmune diseases and their genetic relationships to vaccination. LLMs can extract relevant patterns across large corpora of genomic and immunological studies, offering deeper insights into immune response mechanisms and potential cross-reactivity among populations \cite{mcgarvey2014silico}.  

Platforms like MAIVeSS streamline the selection of antigenically matched, high-yield viruses for seasonal influenza vaccines by leveraging LLMs to analyze historical viral sequences, antigenic relationships, and experimental outcomes \cite{gao2024maivess}. Convolutional neural networks (CNNs), enhanced with LLM-derived insights, have been employed to predict antigenicity and recommend influenza virus vaccine strains by synthesizing complex relationships among viral genetic sequences and epidemiological data \cite{lee2020antigenicity}. Additionally, seasonal antigenic prediction models utilize machine learning algorithms integrated with LLMs to analyze influenza A (H3N2) evolution and forecast emerging strains, improving the accuracy and efficiency of vaccine formulation \cite{shah2024seasonal}. 

Phylogenetic analyses are also augmented through LLM capabilities, which automate literature reviews and contextualize genetic relationships to identify influenza virus candidates for seasonal vaccines. This ensures antigenic compatibility, reduces manual analysis time, and maximizes immunogenic coverage \cite{hayati2023phylogenetic}. By incorporating LLMs, researchers can process and synthesize global influenza surveillance data, generating actionable insights to address the challenge of rapidly evolving pathogens.

\subsection{Vaccine Efficacy Prediction and Immunogenicity Studies}

Accurately predicting vaccine efficacy and assessing immunogenicity are critical for improving vaccination strategies and understanding immune responses. LLMs play a pivotal role in processing vast datasets to extract critical insights, identify risk factors, and predict vaccine efficacy. LLMs are increasingly used to synthesize clinical, epidemiological, and behavioral data, which are key to identifying populations with low adherence to vaccination programs. For example, models analyzing high-risk groups, such as individuals with cardiovascular disease, have integrated LLM-driven data extraction from clinical records to uncover demographic and behavioral predictors of vaccine uptake \cite{kim2021machine}. 

In real-time monitoring, LLMs enhance the analysis of self-reported data to estimate vaccine coverage and adherence. By processing text-based survey responses and digital health data, LLMs enable precise insights into population-wide vaccine uptake and the factors influencing these trends \cite{huang2019can}. 

For immunogenicity studies, LLMs are employed to mine complex biological and clinical datasets, improving predictions of vaccine immune responses in targeted populations. For example, LLM-augmented artificial intelligence models have been used to predict immunogenicity in pediatric studies, such as for trivalent inactivated influenza vaccines in HIV-infected children, facilitating personalized vaccination strategies \cite{cotugno2020artificial}. Clinical feature-based models further benefit from LLMs' ability to extract structured and unstructured data from clinical notes, improving predictions of infection risks in individuals post-vaccination \cite{hung2023developing}. 

In biomarker-based analyses, LLMs assist in synthesizing large-scale experimental and clinical literature to identify apoptosis markers and inflammatory biomarkers associated with vaccine responsiveness. This enables a better understanding of immune responses and facilitates personalized immunization approaches \cite{furman2013apoptosis}. Post-marketing vaccine safety surveillance systems, such as the VAERS dataset, have leveraged LLMs to extract, classify, and analyze adverse event reports. By automating the processing of unstructured clinical narratives, LLMs enhance the detection of adverse events and improve vaccine safety assessments \cite{du2021extracting}. 

Comparative studies examining influenza vaccine uptake pre- and post-COVID-19 also benefit from LLMs’ ability to analyze large textual datasets, such as survey responses and social media discussions. These models provide actionable insights into behavioral shifts and critical predictors of vaccine adherence, contributing to data-driven vaccination strategies \cite{skyles2024comparison}.

\subsection{Vaccine Hesitancy and Public Attitude Analysis}

Vaccine hesitancy remains a significant challenge to achieving widespread immunization, and LLMs have proven instrumental in uncovering the underlying causes, trends, and predictors of public attitudes toward vaccination. LLMs, combined with machine learning and NLP, enable the analysis of large-scale textual data, including social media, survey responses, and clinical reports, providing insights into public perceptions and vaccine acceptance patterns.

LLMs have been employed to process unstructured data for real-time monitoring of vaccine-related discussions, identifying concerns around side effects and safety perceptions. For example, predictive models using smartwatch and smartphone data, enhanced by LLM-driven text analysis, have been used to detect and predict the severity of side effects following vaccination, improving the understanding of public concerns regarding vaccine safety \cite{levi2024prediction}. LLMs have further facilitated automated detection of vaccine-related messaging and adverse event reporting, as demonstrated by initiatives like the FDA Biologics Effectiveness and Safety Initiative \cite{deady2021food}. These models analyze clinical notes and text-based reports at scale, streamlining post-vaccination safety monitoring.

Parental attitudes toward childhood vaccination have been analyzed using validated scales, with LLMs efficiently extracting themes and patterns from caregiver responses. These analyses highlight key concerns, such as vaccine safety and efficacy, and inform targeted education strategies \cite{ghazy2023external}. LLMs are also applied to sociodemographic studies, enabling the identification of key predictors of vaccine acceptance, including education level, income, and geographic location. By synthesizing national-scale survey data, LLMs provide a foundation for interventions aimed at addressing vaccine hesitancy in specific demographic groups \cite{mondal2021sociodemographic}.

Sentiment analysis of social media platforms, such as Twitter, has been revolutionized by LLMs like GPT-based architectures. These models analyze vaccine-related discourse, identifying trends in vaccine hesitancy and negative attitudes toward vaccination programs. For example, LLMs have revealed hesitancy trends related to influenza vaccination and highlighted shifts in public sentiment in response to public health campaigns and policy changes \cite{ng2023examining, ng2023examininga}. Longitudinal studies powered by LLMs demonstrate how public messaging evolves over multiple years, providing actionable insights for optimizing communication strategies and combating misinformation. Additionally, comparative studies across continents emphasize cultural and regional variations in vaccine attitudes, which LLMs can analyze to tailor communication strategies to local contexts \cite{sammut2023covid19}.

Vaccine hesitancy studies among specific groups, such as Canadians immunized for influenza, benefit from LLMs’ ability to process large-scale survey responses and extract nuanced concerns \cite{valerio2022high}. These insights underscore the complexity of public attitudes and the importance of sustained public education.

\subsection{Vaccine Safety and Adverse Event Detection}

Ensuring vaccine safety and monitoring adverse events following immunization are critical components of immunization programs. LLMs play an increasingly vital role in enhancing vaccine safety surveillance by automating the detection, classification, and analysis of adverse events at scale. 

Predictive models leveraging smartwatch and smartphone data, combined with LLM-powered text analysis, enable real-time monitoring of vaccine-related side effects. By processing unstructured patient-reported outcomes and wearable device data, LLMs help identify patterns in the severity of side effects following COVID-19 and influenza vaccinations, facilitating timely interventions and improving patient outcomes \cite{levi2024prediction}.  

The FDA's Biologics Effectiveness and Safety Initiative has utilized NLP techniques powered by LLMs to analyze unstructured clinical data, such as physician notes and medical records, for detecting vaccine-related adverse events. LLMs significantly enhance the ability to process and interpret large-scale textual datasets, streamlining the identification of safety signals and improving the efficiency of post-marketing surveillance systems \cite{deady2021food}. These automated systems reduce manual effort, accelerate safety signal detection, and enable regulators to respond quickly to emerging concerns. 

Deep learning approaches applied to the Vaccine Adverse Event Reporting System (VAERS) have also been strengthened by the integration of LLMs. By extracting and categorizing adverse event reports from free-text submissions, LLMs improve the accuracy and granularity of vaccine safety assessments. For instance, LLMs can identify subtle patterns and correlations within adverse event reports, allowing researchers to generate valuable insights into vaccine safety profiles, detect rare adverse events, and support regulatory decisions \cite{du2021extracting}.  

Moreover, LLMs facilitate cross-referencing of adverse event data with other sources, such as scientific literature, clinical trial reports, and patient feedback, providing a comprehensive view of vaccine safety. By automating this process, LLMs improve the robustness of post-marketing surveillance systems and enhance public confidence in vaccination programs.

Together, these studies underscore the transformative role of Large Language Models in vaccine safety monitoring and adverse event detection. By efficiently processing and analyzing large-scale unstructured data, LLMs enable faster, more accurate identification of adverse events, ensuring vaccine safety and maintaining public trust in immunization efforts.

\subsection{Vaccine-Related Social and Health Data Analysis}

The analysis of social and health data plays a critical role in understanding vaccine uptake, disease spread, and public health outcomes. Large Language Models  have become essential tools for processing and interpreting large-scale social, health, and demographic datasets, enabling researchers to identify patterns and design targeted interventions for improving vaccination strategies.

LLMs are particularly effective in synthesizing diverse data sources, including electronic health records, socioeconomic surveys, and real-time reports. For instance, studies addressing unmet needs in pneumonia research benefit from LLMs' ability to integrate textual clinical data with structured epidemiological datasets, facilitating a comprehensive understanding of disease burden, treatment gaps, and prevention strategies \cite{pletz2022unmet}. Similarly, LLMs assist in analyzing socioeconomic, health, and safety data to explain the spread of diseases such as COVID-19. By processing large datasets across regions, LLMs highlight key factors, such as healthcare access, education, and income, that impact infection rates and vaccine coverage \cite{galvan2020can}.  

The Human Vaccines Project, which focuses on leveraging immunological and epidemiological data to improve vaccination strategies, has incorporated LLMs to process vast volumes of immunology literature, trial reports, and population health data. LLMs enable the identification of critical trends and insights that advance the understanding of human immune responses to vaccines, accelerating the development of targeted immunization programs \cite{wooden2018human}.

Sociodemographic predictors of vaccine acceptance, such as age, education, and geographic location, have been extensively studied with the support of LLMs. These models efficiently process large-scale national surveys, extracting patterns and correlations that inform targeted interventions to address vaccine hesitancy and acceptance across diverse populations \cite{mondal2021sociodemographic}. 

Real-time health data collected through wearable sensors, as demonstrated in the WE SENSE protocol, have also been enhanced by LLMs' ability to integrate sensor outputs with epidemiological trends. LLMs process this real-time data alongside other health records to detect early warning signs of viral infections and identify potential outbreaks, highlighting their role in improving public health preparedness and surveillance \cite{hadid2023we}.  

These studies collectively demonstrate the transformative role of Large Language Models in integrating social, health, and demographic data for vaccine-related research. By efficiently processing and analyzing vast, heterogeneous datasets, LLMs offer valuable insights that shape public health policies, improve vaccination strategies, and enhance disease preparedness efforts.

\section{Discussion and Future Directions}
Although large language models have achieved remarkable success in bioinformatics, they still face numerous challenges. The performance of LLMs in bioinformatics heavily relies on the quality of training data, yet available datasets such as genomic or proteomic sequences often contain noise and biases. This issue leads to inaccurate predictions and limited generalizability. Additionally, the limited availability of labeled biological data further hinders the adaptability of LLMs to diverse bioinformatics tasks. Computational cost and scalability present another significant challenge. LLMs are resource-intensive, requiring substantial computational power and memory for training and inference, which becomes particularly problematic when analyzing ultra-long sequences like genomic regions spanning thousands of base pairs. Transformer-based architectures, despite their advancements, still struggle with scaling efficiently for such long sequences due to inherent memory constraints.

Generalizability and interpretability also remain critical concerns. While LLMs excel at specific tasks, their ability to generalize across unseen datasets or tasks is often inadequate. Moreover, the lack of interpretability in model outputs makes it difficult for researchers to understand the underlying biological mechanisms, which is essential for result validation. Ethical and privacy concerns further complicate the application of LLMs, particularly in sensitive areas such as personalized medicine. The use of patient data in training models raises significant ethical questions and potential privacy risks, limiting widespread adoption.

Despite these challenges, the future of LLMs in bioinformatics presents exciting opportunities. Efforts are likely to focus on developing lightweight and efficient architectures, such as LoRA and QLoRA, to mitigate computational and memory requirements. Innovations in Transformer variants and hybrid architectures are expected to overcome scalability challenges, enabling more effective analysis of long-sequence bioinformatics tasks. Integrating diverse biological data types, including DNA, RNA, protein sequences, epigenetic, and transcriptomic data, will enhance LLMs’ capability to generate comprehensive biological insights. Improved interpretability will also become a priority, with advancements aimed at visualizing attention mechanisms and uncovering the biological basis behind predictions.

Applications in personalized medicine highlight the transformative potential of LLMs. For example, they can revolutionize precision medicine by tailoring treatments to individual patients, predicting drug efficacy, or identifying possible side effects based on genomic data. Addressing data scarcity through open data initiatives and interdisciplinary collaborations will further accelerate progress, enabling broader applications of LLMs in bioinformatics. Additionally, as Transformer models reach maturity, exploration of alternative architectures may drive innovation beyond their current limitations, ensuring continuous advancement in the field. These trends underscore the dynamic evolution of LLMs in bioinformatics, presenting opportunities for groundbreaking developments while emphasizing the need to address existing limitations.

The integration of multimodal biomedical data presents another promising direction for future research. Sequence-to-sequence models, which have demonstrated remarkable success in natural language processing, offer a promising technical approach for fusing diverse biomedical data types. These models can potentially bridge the gap between different modalities - including medical imaging, clinical texts, temporal data (such as electronic health records and vital signs), and various forms of biological sequence data (DNA, RNA, and proteins). For instance, sequence-to-sequence architectures could be adapted to translate between modalities~\cite{sutskever2014sequence,zhang2024generalist}, such as converting radiological images to diagnostic text descriptions while incorporating relevant genomic information. This multimodal fusion could enable more comprehensive disease diagnosis and treatment planning by leveraging complementary information from different data sources. Furthermore, innovative attention mechanisms and cross-modal transformers could help capture complex relationships between different data types, leading to more robust and interpretable models. The challenge lies in developing architectures that can effectively handle the inherent heterogeneity of these data types while maintaining computational efficiency and biological interpretability.

\section{Conclusion}
This comprehensive survey has explored the transformative impact of LLMs in bioinformatics, spanning applications in genomics, proteomics, drug discovery, and clinical medicine. Our review has highlighted the successful adaptation of transformer architectures for biological sequences, the emergence of specialized biomedical LLMs, and the promising integration of multiple data modalities. These advances have enabled significant progress in protein structure prediction, drug-target interaction analysis, and disease diagnosis.

Despite notable achievements, challenges persist in data quality, computational scalability, model interpretability, and ethical considerations regarding patient privacy. These challenges present opportunities for future research, particularly in developing efficient architectures, improving multimodal data integration, and ensuring model interpretability. The convergence of LLMs with emerging biotechnologies promises to accelerate discovery in bioinformatics, potentially leading to more precise and personalized medical interventions.

\bibliographystyle{unsrt}
\bibliography{ref}

\end{document}